# Topology of multiple cross-linked Su-Schrieffer-Heeger chains

A. Sivan[*] and M. Orenstein

*Andrew and Erna Viterbi Faculty of Electrical and Computer Engineering, Technion—Israel Institute of Technology, Technion City, Haifa, 3200003, Israel*

In polymer science, cross-linking of polymer chains yields a substantially modified system compared to the one-dimensional constituent chains, due to the increase of dimensionality and effective seeding by defects (cross-linking sites). Inspired by this concept, we analyze topological features of a unit cell of a generalized topological mesh comprised of several one-dimensional Su-Schrieffer-Heeger (SSH) lattices cross-linked via a single site. The coupling site functions as a defect with protected states in the trivial regime and also induces edges inside the bulk with protected localized states centered around it in the topological regime. When more than two lattices are coupled by the defect, namely, a graph-dimensionality larger than one, the crossed chains support two types of localized eigenstates around the defect. One type is highly controllable by modifying the cross-linking strength, enabling broad tuning of eigenenergies from being submerged in the bulk band to becoming highly isolated and protected. We show that these unique features can be explained by an equivalence of the aforementioned system to an SSH chain coupled nonreciprocally to an external reservoir, yielding a unique pseudospectrum for both the bulk and localized states, with spatially symmetric eigenstates. Applying non-Hermiticity by adding gain and loss to alternating sites, relevant for example to a possible realization of topological coupled-laser fabric, we observe an abrupt transition of the topologically protected mid-gap state from anti-localized to localized near the defect. By changing the gain-loss parameters, we observe a cascaded spatial symmetry breaking of the supported states at exceptional points where parity-time symmetry is broken, for both the localized and the bulk states, exhibiting various novel phases.

## I. INTRODUCTION

Topology of structures and states has applications in many fields in physics. Originating from the concepts of topological insulators in solid-state physics and the quantum hall-effect [1,2], it has been shown that the idea of bulk-edge correspondence and its relation to parity-time ($\mathcal{PT}$) symmetries [3] are also applicable to photonics [4-6]. Various structures that exhibit interesting topological phases have attracted attention, including the Su-Schrieffer-Hegger (SSH) model [7], the Aubry-Andre-Harper (AAH) model [8-10] and the Haldane model [11] to name only few examples. Much research has also been done on realization of topological effects in optical or photonic elements [12-18], with recent experiments demonstrating topological insulator lasers [19,20] and recently also coherent lasing along the interface between two-dimensional topological arrays of vertical-cavity surface-emitting lasers (VCSEL) [21].

Various extensions of lattice models and their topological characteristics have been studied. Non-Hermiticity was introduced to the SSH lattice by addition of on-site gain and loss, or by imaginary and/or non-reciprocal couplings between sites, which resulted in richer physical phenomena including generalized topological phases attributed to the $\mathcal{PT}$-symmetry operator [14,15,22-32], non-Hermitian skin-effect [33-38], and more. Research of the topological effects and phases of an SSH lattice in two-dimensions was also conducted, generalizing the one-dimensional (1D) bulk-edge correspondence to higher dimensions and demonstrating the emergence of higher-order protected edges [39-46]. Other extensions with non-local couplings and other symmetries were also reported [47-51].

The effects of defects and discontinuities on the states and dynamics of SSH lattices have been considered in literature. It has been suggested that the incorporation of either one or more sites that differ from the embedding lattice, or on the interface between distinct lattices, introduces states with unique characteristics such as localization and topology-induced robustness against noise [12,13,52-62]. These defects enrich the dynamics of the pure lattices and can be used to tailor desired eigenspectra and eigenstates. However, to our knowledge, these works are reported primarily in a 1D framework, and defects in higher-dimensional systems were not rigorously studied. This is a considerable gap in the understanding of the effects lattice defects have on realistic topological arrays.

In this work, we study the influence of defects on the dynamics of lattices with a dimension higher than one. A plausible scheme for gradually increasing the dimensionality is to form a crosshatched network of SSH chains coupled at each intersection point, while reducing the distance between those intersections. Here we present a rigorous analysis of a building block of such a network, consisting of several crossed SSH lattices coupled by a single mutual defect site in both the Hermitian and non-Hermitian regimes. The model is general to any number of similar or dissimilar lattices, and in this paper, we focus for concreteness on the simple crossed-chain structure – a structure consisting of four identical 1D SSH lattices connected to a mutual coupling element. This configuration is novel in the sense that it spans more than one dimension (as will be further explained in section II), and the defect site acts as a source for the generation of induced edges in the mutual boundaries of the four 1D topological lattices. This produces several key results. In the trivial regime, we find two different types of localized states that are supported by this structure, while in 1D only one type is resolved, unless a more complex defect is included (see section III). We denote those two types as the defect (induced) states and the zero-energy state. The former has energies residing outside of the bulk pseudospectrum while the latter is a midgap state with a zero energy as imposed by symmetry considerations. In the topological regime the zero-energy state becomes anti-localized, and a third type of localized states exists, which is the well-known topological edge states of free SSH chains. We show that the crossed-chain system can be described by three identical disconnected SSH lattices, and a fourth SSH lattice coupled to an effective external reservoir via a non-reciprocal coupling. The latter constitutes an energy ladder shifted from that of a free SSH chain for both the localized and bulk states that are associated with the defect coupling. Those states are shown to be spatially symmetric. The system is then extended to a non-Hermitian setup by adding on-site imaginary potentials in a symmetry-preserving scheme, describing gain and loss. We find that although gain and loss were added, the defect-associated states have the same wavenumbers as in the Hermitian case. Furthermore, we show how the spatial symmetries of the amplitudes of the eigenstates break in a sequence of exceptional points. Our proposed framework therefore expands the understanding of the influence of defects in higher dimensions and suggests that they can be exploited to control the dynamics of a higher-dimensional system. As mentioned, the crossed-chain structure can be used as a building block to create a fully 2D protected state by effectively creating an array of internal edges within the lattice bulk, rather than the more classical 1D protected edge states along the boundary walls between different 2D lattices.

---

[*] amirsi@campus.technion.ac.il

The paper is organized as follows. In Section II we present the Hermitian crossed-chain SSH structure, and emphasize the dynamics introduced by the higher-dimensionality of the system. In Section III we rigorously analyze the effect of the defect coupling strength on the energy pseudospectrum of the structure. In Section IV we generalize our framework by adding non-Hermiticity to the system. We conclude and further discuss the results in Section V.

## II. THE HERMITIAN CROSSED-CHAIN SSH CONFIGURATION

The Hermitian crossed-chain configuration is comprised of $K$ finite SSH chains connected to a mutual additional lattice element as depicted in FIG. 1(a) for the particular case of $K=4$ identical chains. We will henceforth refer to this additional element as the defect site. In this section we explore the case of a "basic" defect that we will soon define. Scattering-theory methods such as Schwinger-Lippmann may be useful when considering simple lattices and some extensions [63-69], but are generally intractable when schemes that are more complicated are considered. We will therefore solve the Hamiltonian eigenvalue problem directly.

The Hamiltonian of a single Hermitian SSH chain indexed by $\sigma$ and formed by $M_\sigma \in 2\mathbb{N}$ sites with a unit distance between them, is

$$H_\sigma = \sum_j^{M_\sigma - 1} a_j^\sigma \phi_j^{\sigma\dagger} \phi_{j+1}^\sigma + H.c. \qquad (1)$$

$$a_j^\sigma = \begin{cases} a^\sigma & \forall j \in 2\mathbb{N} \\ \tilde{a}^\sigma & \forall j \in 2\mathbb{N}+1 \end{cases} \qquad (2)$$

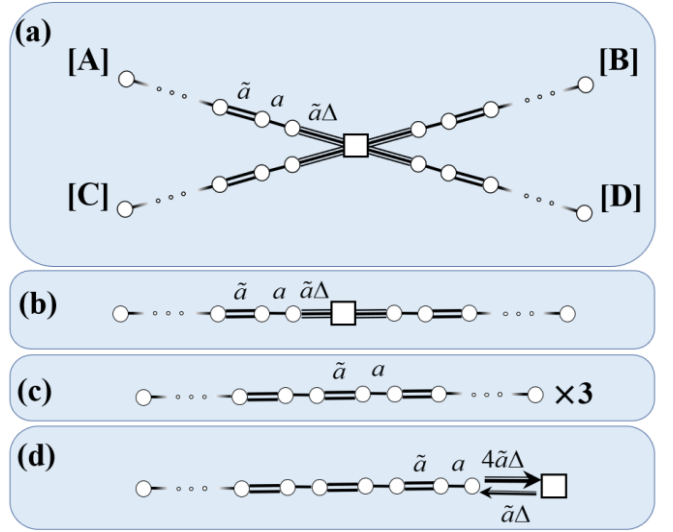

FIG. 1. (a) Schematic illustration of the symmetric crossed-chain configuration. Black lines denote a coupling strength $a$, double-lines a coupling strength $\tilde{a}$ and triple-lines the defect coupling strength $\tilde{a}\Delta$. (b) 1D mirrored SSH chain structure, that was analyzed in e.g. [16]. The configuration of (a) is isospectral with (c) + (d); (c) The degenerate triplet eigenvalues are equivalent to those of three detached SSH chains of the same length and coupling coefficients as these of its constituent chains, and (d) the eigenvalues of the singlet eigenvalues are equivalent to those of the same SSH chain connected to an additional element via a non-reciprocal coupling strength.

Here $\phi_j^\sigma$ are the amplitudes at the $j$'th site, H.c denotes the Hermitian conjugate, and $a^\sigma$, $\tilde{a}^\sigma$ are the respective intra- and inter-cell coupling coefficients that are assumed to be real and positive. The crossed-chain SSH structure (FIG. 1(a)) is an intersection of four SSH chains of lengths $M_\sigma$ with $\sigma \in \{A,B,C,D\}$ as defined by Eq. (1), through a connection to a common site described by a defect site term $U$, such that

$$H_{tot} = U + \sum_\sigma \sum_j^{M_\sigma - 1} a_j^\sigma \phi_j^{\sigma\dagger} \phi_{j+1}^\sigma + H.c , \qquad (3)$$

$$U = \sum_\sigma \tilde{a}^\sigma \Delta^\sigma \phi_{M_\sigma}^{\sigma\dagger} \phi_\Delta + H.c. \qquad (4)$$

where the coupling strength of the defect to the end of the $\sigma$ SSH chain is $\tilde{a}^\sigma \Delta^\sigma$ and the defect site amplitude is $\phi_\Delta$. In contrast to a 1D system consisting of two SSH chains around a defect in a mirrored setting (FIG. 1(b)), that was treated in previous works (e.g. [12,13,16]) – the crossed-chain structure has more than one dimension. The measure of dimensionality in our case is borrowed from graph theory: a star graph (the defect is the central node and the SSH chains are the leaves) is 1D for $K \leq 2$ and 2D for $K > 2$. This definition is different from geometrical two-dimensionality (e.g. the 2D SSH lattice in [39]).

We define a "basic defect" as a site that acts as a boundary wall between the $K$ SSH lattices, with $\Delta^\sigma = \Delta = 1$ for every $\sigma$. This defect is basic in the sense that it only acts as a boundary wall between several lattices, but it does not modify the coupling scheme of any of them. Systems with a non-basic defect ($\Delta \neq 1$) will be addressed in the next sections. Furthermore, from this point on, we will discuss only "isotropic" systems with $K = 4$, wherein $M^\sigma = M$, $a^\sigma = a$ and $\tilde{a}^\sigma = \tilde{a}$ for all $\sigma$.

The increase in dimensionality introduces new features that are absent from the 1D case.

The first feature, occurring in the topologically-trivial regime, is that additional localized states are now supported by the system. While the trivial 1D structure (for which $K = 2$) contains one localized defect state, the trivial crossed-chain structure has three non-degenerate localized defect states. In FIG. 2(a), we illustrate the pseudospectrum of a crossed-chain structure in the trivial regime, with insets focusing on the localized defect states depicted in FIG. 2(c)(i)-(iii). One of the localized defect states has zero energy with a localization length of $\eta_0 = -\log(\tilde{a}/a)$ and is localized on the defect (FIG. 2(c)(ii)), as in the 1D case [12], while two novel localized defect states appear with energies given by

$$E_\Delta^2 = a^2 + \tilde{a}^2 + 2a\tilde{a}\cosh\kappa, \qquad (5)$$

where the localization length $\kappa \in \mathbb{R}^+$ is related to the parameters of the structure by

$$K = 1 + \frac{a}{\tilde{a}} e^\kappa \qquad (6)$$

(Appendix A). These two states are also localized on the defect site itself. Eq. (6) has a solution for $\kappa$ in the trivial regime $\tilde{a} < a$ if and only if $K > 2$. Therefore, while the zero-energy defect state (henceforth denoted as $E_0$) stems from the symmetry of the structure and also exists in the 1D case, the emergence of these two additional localized defect states is a feature attributed to the higher dimensionality of the system.

The second important feature of the higher-dimensionality occurs in the topological regime $\tilde{a} > a$ and is the distinction of the regular edge states from the localized defect states and zero-energy state, in terms of their localization lengths.

The topological regime here is obtained under the same coupling relations as for the isolated SSH chain ($\tilde{a} > a$). Although rigorous calculation of the topological invariant (e.g., Chern number) of our aperiodic structure is a formidable task, the transition from a trivial to a topological phase is clear and exhibits the typical edge-bulk correspondence and chiral symmetry breaking at the bandgap closing [70,71]. The chiral symmetry is consistent with the relation $\sigma_z H_{tot} \sigma_z = -H_{tot}$ that is fulfilled for the matrix $\sigma_z$ generalized from its definition in [16], and is indicated by the fact that the eigenvalues of the system come in $(-E_m, E_m)$ pairs, as we explicitly show in the next section.

In the topological regime we get two localized defect states $\pm E_\Delta$ and seven other localized edge states inside the bulk gap (FIG. 2(b)). The seven localized states (FIG. 2(d)(ii)-(viii)) are further separated to one zero-energy state $E_0$ (FIG. 2(d)(v)) and six other edge states $E_E$. Similar states were briefly mentioned in analyses of 1D systems in the topological regime [72,73], however, even for the 1D system no rigorous exploration has been conducted to the best of our knowledge, and thus we provide a concise derivation in Appendix E. The zero-energy state $E_0$ is a symmetric state with a localization length $\eta_0 = |\log(a/\tilde{a})|$, anti-localized with respect to the defect site. The $E_E$ states are showing mixed localization – both on the outer edges and on the emerging induced edges neighboring with the defect site. These states exhibit exponential localization only in the long-lattice limit, with localization length $\eta_E \approx -\log(a/\tilde{a})$ [74]. In other words, while the energy of $E_0$ is identically zero and its amplitude profile is exactly exponential for all lengths and lattice coefficients, the energies of the states $E_E$ only approach zero and an exponential localization for a long lattice. In this paper we employ system parameters for which the long lattice approximation $\eta_E \approx -\log(a/\tilde{a})$ holds.

When solving Eq. (6) in the topological regime, one can see that $K = 2$ yields $\kappa = -\log(a/\tilde{a})$, which means that the localization lengths of the localized defect state and of the topological edge states are equal in 1D. For a higher-dimensional system ($K > 2$), the localized defect states $\pm E_\Delta$ have different localization lengths from those of the edge states. This result indicates that in a higher-dimensional system, the state localization due to the topological nature of the SSH chains is a distinct physical phenomenon from the state localization due to the presence of the defect.

We have thus shown that in the crossed SSH trivial regime, non-mid-gap localized states emerge as a result of the effective (internal) edge induced by the coupling defect. This type of states does not exist in the previously researched $K = 2$ structure in the trivial case. We have thus observed three different types of localized states – the zero-energy state $E_0$ and the localized defect states $\pm E_\Delta$ that originate from the defect – both of which exist also in the trivial regime, and in addition – the edge states $E_E$ that are due to the topology of the SSH chains in the topological regime.

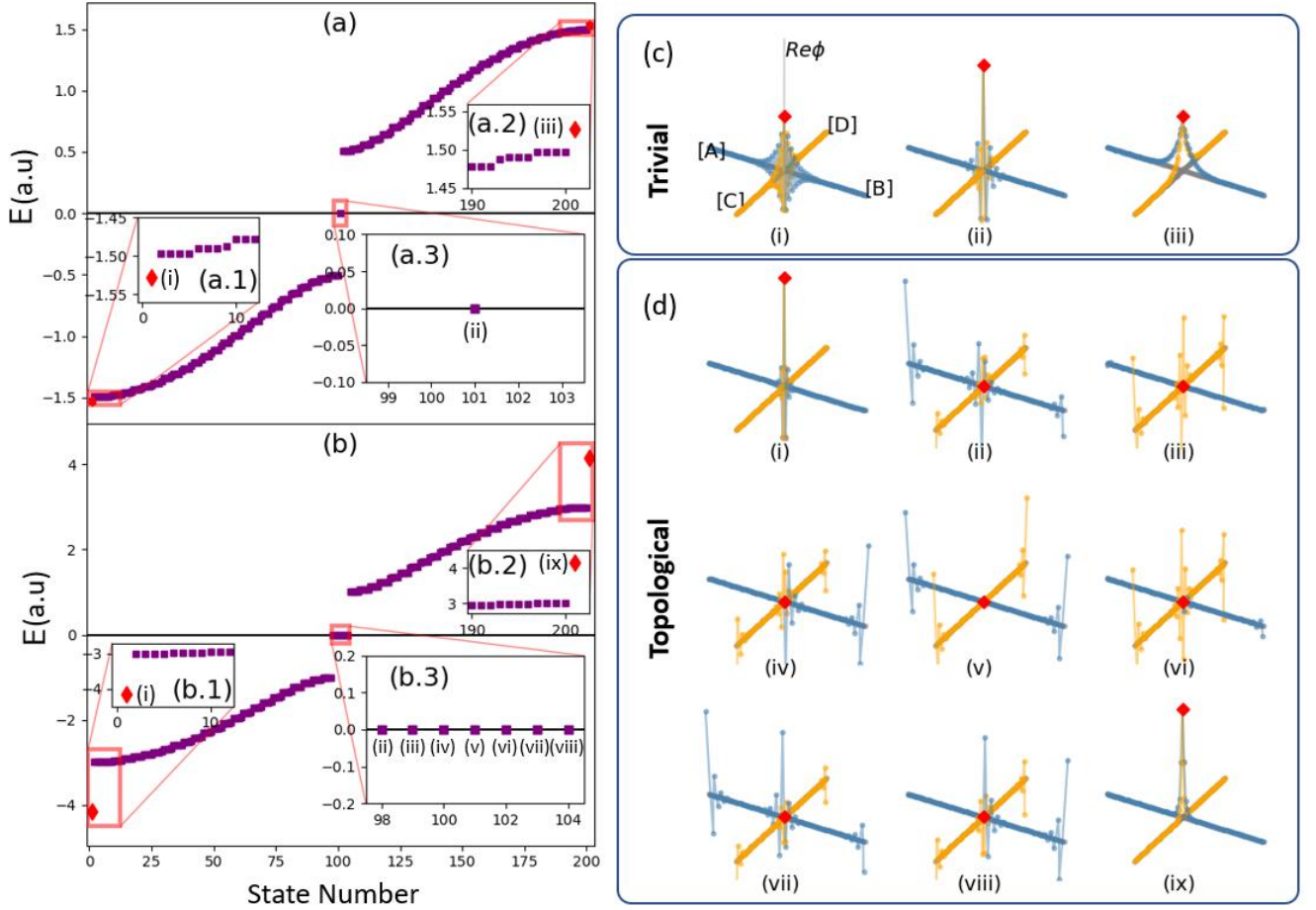

FIG. 2. Energy pseudo-bands for symmetric crossed-chain structures with $M = 50$ and $a = 1$. (a) Trivial SSH chains with $\tilde{a} = 1/2$. (b) Topological SSH chains with $\tilde{a} = 2$. The insets (a.)(1-3), (b.)(1-3) are magnified portions of the pseudospectra. Non-mid-gap localized states are marked with red diamonds. Insets (a.3), (b.3) – zero-energy and edge states. (c),(d) Amplitudes of the localized states in the crossed-chain structure, with SSH chains [A] through [D], numbered in an ascending order with respect to energy. (c)(i-iii) The localized states for the trivial case (Fig.2(a)); c(i) and c(iii) are the non-mid-gap defect states, c(ii) is the zero-energy state. (d)(i-ix) The localized states for the topological case (Fig.2(b)); d(i) and d(ix) are the localized defect states, d(v) is the defect zero energy state localized on the outer edges of the structure, d(ii-iv) and d(vi-viii) are the edge states localized on the edges of the constituent SSH chains. Red diamonds mark the defect site amplitudes in (c) and (d).

## III. CONTROLLING PROTECTED STATES BY TUNING OF DEFECT COUPLING

We analyze the effect of modifying the effective defect coupling strength $\Delta$ on the pseudospectrum and eigenstates of the system, generalizing the results of the previous section to more complicated defects. The observation that the coupled cross chain system is isospectral with an equivalent system consisting of two manifolds – one is a "degenerate trivial systems" made of $K-1$ simple uncoupled SSH chains (FIG. 1(c)) and the second is a single SSH lattice coupled to a nonreciprocal edge element (FIG. 1(d)), is a cornerstone in our analysis. This will be further explained and proven. Furthermore, for some values of $\Delta \neq 1$ one can find localized states, with localization lengths that were exhibited in section II only for the 2D case, in the 1D ($K = 2$) settings.

Starting with the extreme case of $\Delta = 0$, the system is comprised of four identical disconnected SSH chains and a single isolated defect site and so the pseudospectrum is quadruply-degenerate plus one. In the trivial regime, there is only one localized state – the zero-energy state residing on the defect site. In the topological regime there are nine localized states – two edge states of each of the identical topological SSH chains, and one zero-energy state completely localized on the defect. The non-zero energies are related to a discrete wavenumber $k_m$ via the well-known SSH lattice dispersion relation $E^2(k_m) = a^2 + \tilde{a}^2 + 2a\tilde{a}\cos k_m$ in both the trivial and topological regimes.

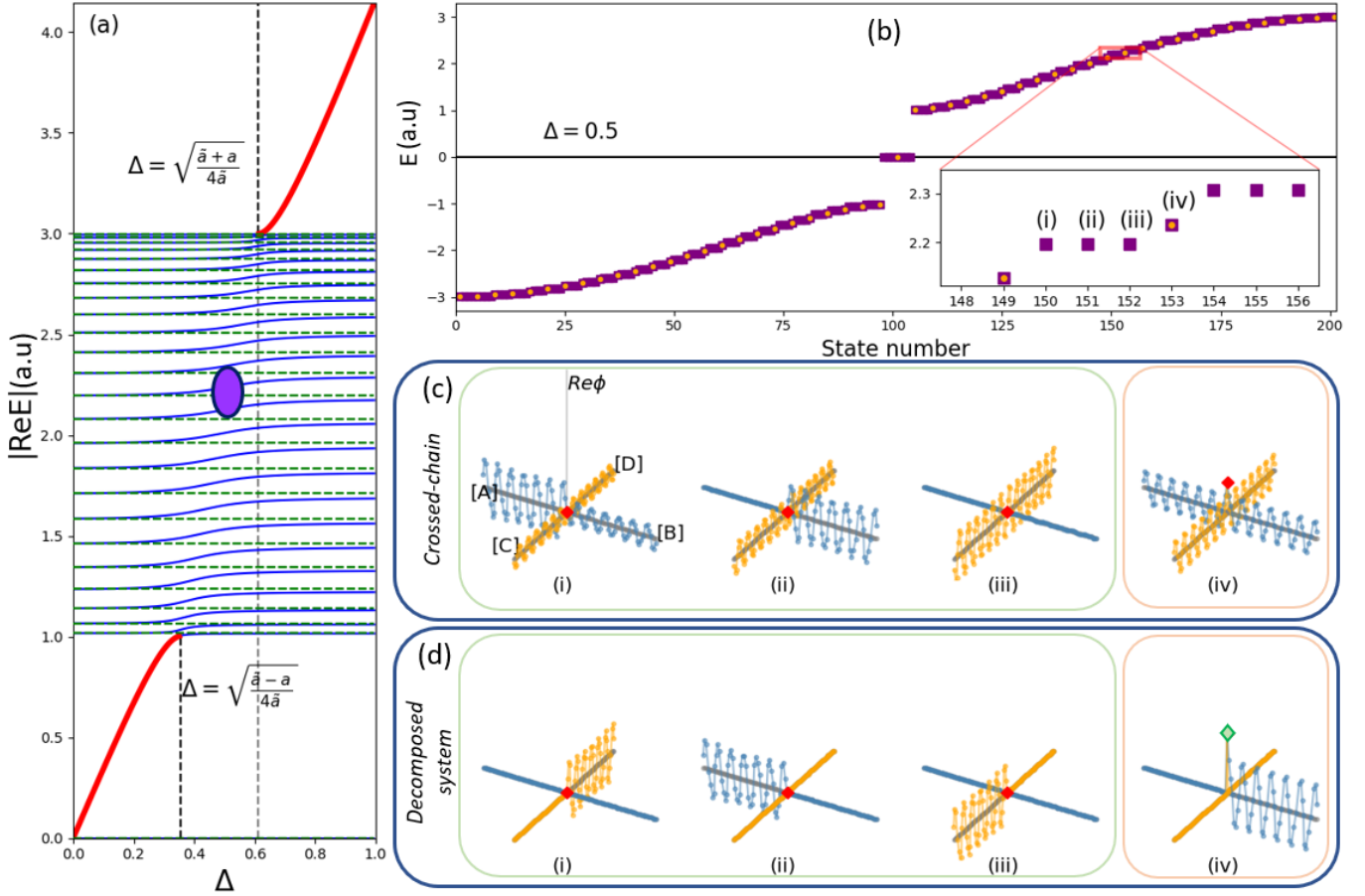

FIG. 3. (a) Positive real-energy pseudospectrum for the symmetric crossed-chain system of FIG. 1(a) in the topological regime with $\tilde{a} = 2a$ and $M = 50$ vs. defect coupling strength coefficient $\Delta$. Dashed green lines denote the energy levels of the triplet pseudospectrum, blue lines denote the energy levels of the singlet pseudospectrum, and the thick red lines denote the isolated localized defect states. Vertical dashed lines mark the critical values of $\Delta$ between which the localized defect states are merged into the bulk. (b) Energy pseudospectrum at $\Delta = 0.5$ (the purple marker in subfigure (a)). The yellow dots mark the singlet pseudospectrum. Inset: an example triplet bulk states (marked i-iii) and the associated singlet state (marked iv). Real values of the amplitudes of the states marked in this inset appear for (c) the crossed-chain system and (d) the reduced representation by an isospectral system comprised of three detached SSH chains and one SSH chain coupled non-reciprocally to a defect site. Defect site amplitude marked by a red diamond, and the amplitude of the non-reciprocally coupled defect is marked by a light-green diamond.

For the general case $\Delta > 0$, the aforementioned decomposition is manifested, as the quadruply-degenerate bulk pseudospectrum with energies $E_m(k_m)$ breaks into two manifolds; a triplet: a triply-degenerate pseudospectrum with energies $E_T(k_m) = E_m(k_m)$ and a singlet: a pseudospectrum with shifted (respective to the free SSH) eigenenergies $E_S(p_m)$ and shifted wavenumbers $p_m \equiv k_m + \delta k_m$ (FIG. 3(a)). The transformation of each degenerate quartet into a degenerate-triplet and a singlet state with shifted energies and wavenumbers, results, as may be expected, from coupling of the SSH chains to the defect.

The bulk energy of the m'th triplet is given by $E_T^2(k_m) = a^2 + \tilde{a}^2 + 2a\tilde{a}\cos k_m$ (Appendix B) which is the energy of the m'th state of an isolated single SSH chain (FIG. 1(c)). In the topological regime, this manifold also includes a triply-degenerate pair of topological edge states $E_E$ on the edges of the chains with localization lengths $\eta_E \approx -\log(a/\tilde{a})$ that do not depend on $\Delta$. In the triplet manifold, since the energy of the bulk states $E_T(k_m)$ is equal to the energy of the bulk states of a free SSH chain $E_m(k_m)$, each m'th triplet state is comprised of three degenerate superpositions of the m'th bulk states of the four SSH chains. Additionally, since the free SSH chains do not include the defect site, the defect site amplitude in the triplet states must be zero. The singlet bulk state shifted from the m'th band must be spatially symmetric with respect to the central element (FIG. 3(b)). Moreover, the energy shift of the singlet bulk is related to a wavenumber shift which, as shown in Appendix B, implies a non-zero amplitude on the defect site together with the symmetrical extended amplitude profile on the chains.

The singlet part of the pseudospectrum is more interesting since it unfolds the impact of the defect and can be obtained from analyzing an equivalent non-Hermitian system. Since the singlet states are symmetric around the defect site, in the reduced representation we consider an effective system comprised of a single SSH chain connected to an external "reservoir" with a coupling constant $\tilde{a}\Delta$ (FIG. 1(d)). However, since the defect site in the crossed-chain structure experiences coupling to four identical adjacent sites, the reservoir in the effective system is coupled back to the effective SSH lattice with a coupling strength of $4\tilde{a}\Delta$. This introduces non-Hermiticity to the Hamiltonian of the effective system, although it still supports only real eigenenergies. The bulk energies of the singlet manifold are given by

$$E_S^2(p_m) = a^2 + a\tilde{a}\frac{\sin[p_m(M/2-1)]}{\sin(p_m M/2)} + 4(\Delta\tilde{a})^2 \quad (7)$$

and the wavenumbers $p_m = k_m + \delta k_m$ are the solutions of an implicit equation derived in Appendix B. It should be noted that since (7) is an even function, its solutions are $(-E_S, E_S)$ pairs. For defect coupling strengths of $\Delta^2 < (\tilde{a}-a)/4\tilde{a}$ and $\Delta^2 > (\tilde{a}+a)/4\tilde{a}$, the localized states emerge from the bulk with energies

$$E_\Delta^2 = a^2 + \tilde{a}^2 - 2a\tilde{a}\cosh\kappa \quad (8)$$

$$E_\Delta^2 = a^2 + \tilde{a}^2 + 2a\tilde{a}\cosh\kappa, \quad (9)$$

respectively (Appendix A). These are illustrated by the red curves in FIG. 3(a).

Unlike the localization lengths $\eta_0$ and $\eta_E$ that do not depend on the defect strength $\Delta$, in the defect-induced states there is a clear dependence given by (Appendix A)

$$\Delta^2 = \frac{\tilde{a} + a\lambda(\kappa)^{-1}}{4\tilde{a}}, \quad (10)$$

where $\lambda(\kappa) = -e^{-\kappa}$ if the localized defect-state energies $\pm E_\Delta$ are within the gap ($|E_\Delta| < \tilde{a} - a$) and $\lambda(\kappa) = e^{-\kappa}$ if they are outside the bulk pseudospectrum ($|E_\Delta| > \tilde{a} + a$). The former exists only in the topological regime. Another state belonging to the singlet mode is trivial solution $E_S = 0$ which is the zero-energy state $E_0$, localized on the defect site in the trivial regime and on the outer edges of the structure in the topological regime. Similar to the case of a basic defect, the zero-energy state $E_0$ has a localization length given by $\eta_0 = |\log(a/\tilde{a})|$, independent of $\Delta$ for $\Delta \neq 0$.

We have shown that the Hermitian crossed-chain system can be fully described by an isospectral effective system that is a combination of a triply-degenerate SSH lattice and a non-Hermitian system consisting of an SSH coupled to an external non-reciprocal reservoir. The energy pseudospectrum can therefore be denoted as $E(k,\Delta) = \{E_0, E_S(p_m,\Delta), 3E_T(k_m)\}$, with $m \in \pm\{1....M/2\}$. We emphasize that this decomposition is exact, and that the reduced representation of the system and original representation of Eq. (3) are spectrally equivalent. $E(k,\Delta)$ is comprised of $(-E_m, E_m)$ pairs and a zero-energy state, indicating the chiral symmetry mentioned in Section II.

This degeneracy breaking into singlet and triplet manifolds, can be easily generalized to the $K$-chains case in which the pseudospectrum splits into a singlet and $(K-1)$-fold-degenerate manifolds. This is observable only in systems of dimensionality higher than one, since in the 1D system ($K=2$), the splitting of a degenerate doublet pseudospectrum is into two distinct singlet manifolds; in one manifold, the extended amplitude distributions are anti-symmetric with wavenumber $k_m$, and in the other are symmetric with wavenumber $k_m + \delta k_m$ and a non-zero amplitude on the defect. Therefore, no degeneracy occurs.

## IV. NON-HERMITIAN CROSSED-CHAINS

### A. Model of the non-Hermitian structure

We generalize the crossed-chain Hamiltonian to a non-Hermitian setting, by introducing imaginary on-site potentials describing loss and gain. For concreteness, we studied the case of an alternating gain-loss pattern along each of the four coupled chains, with a transparent defect. The complex SSH Hamiltonian is

$$H_\sigma = i\sum_j^{M_\sigma} \gamma_j^\sigma \phi_j^{\sigma\dagger}\phi_j^\sigma + \sum_j^{M_\sigma - 1} a_j^\sigma \phi_j^{\sigma\dagger}\phi_{j+1}^\sigma + H.c, \quad (11)$$

with the non-diagonal coefficients generally defined as in Eq. (2) and $\gamma_j^\sigma = (-1)^j \gamma^\sigma$, $\gamma^\sigma \in \mathbb{R}^+$ are the gain-loss coefficients. For characterizing our system, we employ parity ($\mathcal{P}$), time ($\mathcal{T}$) and parity-time ($\mathcal{PT}$) symmetries. While the conventional definition of the time-reversal operator $\mathcal{T}(-i)\mathcal{T}^{-1} = i$ is directly applicable in our system, attempting to generalize the $\mathcal{P}$- and $\mathcal{PT}$-symmetries to the crossed-chain configuration is not trivial; the definition of $\mathcal{P}$-symmetry is not unique for a structure that is not 1D since the system cannot be described by a single spatial axis. Our parity operator is "unbroken" if the system is invariant under permutation of any of the $K$ SSH chains (this coincides with the usual definition $\mathcal{P}x\mathcal{P}^{-1} = M+1-x$ for the case of 1D). We use the notation "$\mathcal{PT}$-symmetry" more loosely for indicating a situation where the eigenenergies are real (quasi-Hermitian operator [3]).

For $\forall \sigma: M_\sigma = M$, $a^\sigma = a$, $\tilde{a}^\sigma = \tilde{a}$, $\gamma^\sigma = \gamma$, as demonstrated in FIG. 4, the crossed-chain system can be analyzed either as two SSH systems, each maintaining $\mathcal{PT}$-symmetry (but not $\mathcal{P}$-symmetry) coupled by a defect, or as two SSH systems, each preserving $\mathcal{P}$-symmetry but breaking $\mathcal{PT}$-symmetry, coupled by the same defect. This ambivalence stems from the dimensionality imposed by the coupling through the defect site, and has no 1D equivalence.

We denote without loss of generality $\gamma_j^{A,C} = (-1)^j \gamma^{A,C}$ and $\gamma_j^{B,D} = -\gamma_j^{A,C}$. Re-writing Eq. (3) c.f. Eq. (4) to include the non-Hermiticity yields,

$$H_{tot,\gamma} = \sum_\sigma \tilde{a}^\sigma \Delta^\sigma \phi_{M_\sigma}^{\sigma\dagger} \phi_\Delta + \sum_\sigma \sum_j^{M_\sigma} (-1)^j i\gamma^\sigma \phi_j^{\sigma\dagger} \phi_j^\sigma \\ + \sum_\sigma \sum_j^{M_\sigma - 1} a_j^\sigma \phi_j^{\sigma\dagger} \phi_{j+1}^\sigma + H.c. \quad (12)$$

As in the previous sections, we will consider the "isotropic" case $\forall \sigma: M_\sigma = M$, $a^\sigma = a$, $\tilde{a}^\sigma = \tilde{a}$, $\gamma^{B/D} = \gamma$.

In general – topological protected states are expected to exhibit enhanced robustness to disorder of the lattice. This is a good indicator for a topological phase, especially for the non-Hermitian case where the other indicators such as topological invariants and bandgap closing are not well defined. We performed a thorough robustness analysis for all "protected" states in the topological phase of the non-Hermitian case under chirality- and $\mathcal{PT}$-symmetry-preserving disorder (as was done before for similar cases, e.g. in [13,16]). The analysis detailed in Appendix G, shows that even for disorder parameters of $\sim 1 - \gamma/|\tilde{a} - a|$ (the maximal disorder of given parameters above which $\mathcal{PT}$ symmetry may be broken, as explained in the appendix), the variances of the distributions of numerically calculated wavenumbers for the topological states in hundreds of realizations amount to less than 1.6% of their respective mean values. This demonstrates substantial robustness for the zero states, edge states and defect states, thereby validating the robustness of the non-Hermitian topological phase.

Unlike the Hermitian case – when the non-Hermiticity is introduced – the system is comprised of two identical pairs of complex-conjugate SSH chains, rather than four identical SSH chains. Therefore, there are two possible choices of effective systems and SSH chains, that are identical up to a complex-conjugation operation (Appendix D). Thus, although the amplitude profiles of the states are not identical on the four SSH chains, knowledge of the profile of one chain in a singlet state suffices to unambiguously describe the entire system. Moreover, since the full crossed-chain system and the reduced equivalent systems are all $\mathcal{T}$-symmetric, complex conjugation does not affect their pseudospectra. Thus, the effective system would yield valid isospectral representations of the system and we can use the equivalent system of one non-Hermitian SSH chain with non-reciprocal coupling to a reservoir and three non-Hermitian SSH chains.

The pseudospectrum of the system is $E(k,\Delta,\gamma) = \{E_0, E_{S,\gamma}(p_m,\Delta), 3E_{T,\gamma}(k_m)\}$ for $m \in \pm\{1...M/2\}$ where $E_{S,\gamma}(p_m,\Delta)$ and $E_{T,\gamma}(k_m)$ are the gain-parameter-dependent singlet and triplet energies, respectively. It is interesting to note that the shifts in the wavenumbers of the bulk states associated with the singlet state are exactly identical to those obtained for the Hermitian case, and do not depend on $\gamma$ (Appendix D), making the result of the previous section much more general.

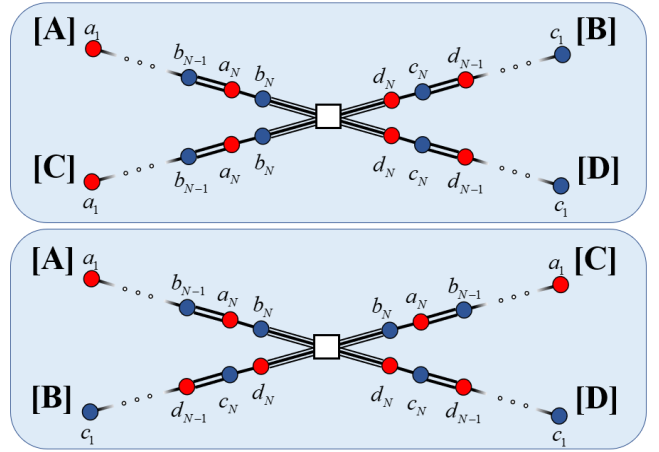

FIG. 4. Illustration of the symmetric non-Hermitian crossed-chain $\mathcal{T}$-symmetric configuration. Red sites denote gain $\gamma$ and blue circles denote loss $-\gamma$. The two illustrations portray an identical system; the crossed-chain scheme can be analyzed as the systems comprised of chains $\{A,B\}$ and $\{C,D\}$, or chains $\{A,C\}$ and $\{B,D\}$, coupled through the defect site.

The bulk energies of the non-Hermitian system are given by

$$E_{T,\gamma}^2(k_m) = a^2 + \tilde{a}^2 + 2a\tilde{a}\cos k_m - \gamma^2, \quad (13)$$

$$E_{S,\gamma}^2(p_m,\Delta) = a^2 + \tilde{a}^2 + 2a\tilde{a}\cos p_m - \gamma^2, \quad (14)$$

for the non-Hermitian triplet and singlet manifolds, respectively, where $k_m$ are the discrete wavenumbers of a constituent SSH chain with $\gamma = 0$, and $p_m$ are the discrete solutions of the implicit function as described in Section III. The energies of the localized states are generalized to include the non-Hermiticity parameter $\gamma$ (Appendix C). The defect states of the singlet manifold are

$$E_{\Delta,\gamma}^2 = a^2 + \tilde{a}^2 \pm 2a\tilde{a}\cosh\kappa(\Delta) - \gamma^2, \quad (15)$$

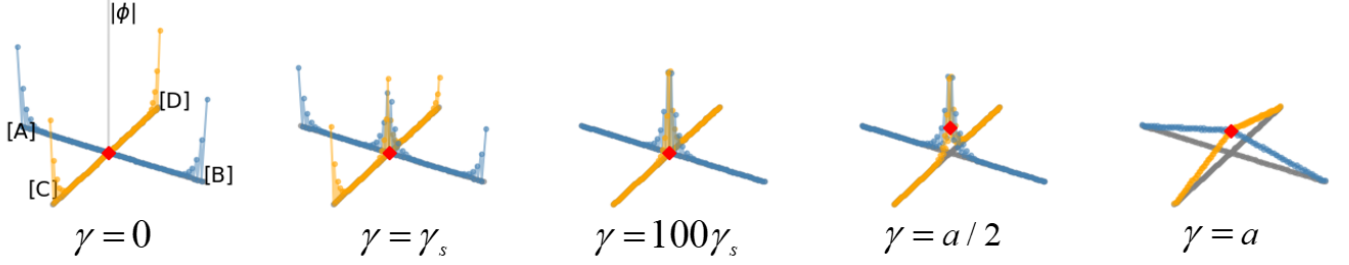

FIG. 5. The zero-energy state in the topological regime for different values of $\gamma$ with SSH parameters $a=1$, $\tilde{a}=2$ and $M=50$. The transition from outer to inner edge localization occurs under these conditions for $\gamma_s \approx 4.4 \times 10^{-8}$, which is virtually zero. The transition is even more abrupt when further increasing the chain lengths.

while the triplet topological edge states (in the long lattice limit) are

$$E_{E,\gamma}^2 = a^2 + \tilde{a}^2 - 2a\tilde{a}\cosh\eta_E - \gamma^2. \quad (16)$$

We show in Appendix C that the localization length $\kappa$ of the localized defect states $\pm E_\Delta$ does not depend on $\gamma$, but depends on $\Delta$ exactly as in the Hermitian case (see Eq. (10)). The edge states (16) exist of course only in the topological regime. In the Hermitian case the edge localization length was given by $\eta_E \approx -\log(a/\tilde{a})$, and it can be readily obtained from the Hamiltonian equations that $\eta_E$ does not depend on $\gamma$. Substituting $\eta_E$ in Eq. (16), we obtain that the energies of the edge states are $E_{E,\gamma} \approx \pm i\gamma$, which are purely imaginary.

We emphasize that the distinction between the localization constants of the edge states and defect states for a basic defect is a property of the higher-dimensionality of the system, also when non-Hermiticity is introduced.

The third type of a localized state is the zero-energy state $E_0$ discussed in the previous sections. Its localization length is obtained from eliminating the LHS of the non-Hermitian extension of the zero-energy state dispersion relation [16],

$$\eta_0 = \eta_0(\gamma) = \text{arc}\cosh\left(\frac{a^2 + \tilde{a}^2 - \gamma^2}{2a\tilde{a}}\right). \quad (17)$$

Recall that in the Hermitian case we obtain $\eta_E \approx \eta_0(\gamma=0) = -\log(a/\tilde{a})$ and $E_E \approx E_0 = 0$ in the long-lattice approximation. Therefore, an important signature of the non-Hermiticity in the topological regime is the further generation of a distinction between the zero-energy state $E_0$ and the triply-degenerate topological SSH edge states $E_E$.

We may wonder about the discontinuous characteristics of the localization, between the Hermitian and the non-Hermitian cases, when the addition of very small gain-loss ($\gamma \to 0^+$) prompts a very abrupt change in the spatial localization of the zero-energy state from the outer edges to the inner (defect induced) edges (FIG. 5). This abrupt transition is a consequence of the long-chain approximation; without this approximation, $E_E$ is small but finite except at a specific small value of $\gamma = \gamma_s$ for which $E_E(\gamma_s) = 0$ and then the edge states are exactly degenerate with the zero-energy state. For $0 < \gamma < \gamma_s$, the zero-energy state is localized on the far edges of the SSH chains, but as $\gamma$ is increased, localizations appear on the inner edges and increase in magnitude. At $\gamma = \gamma_s$, the amplitudes at the induced and outer SSH edges are equal up to a phase, and for $\gamma > \gamma_s$ the amplitudes become localized only on the induced edges around the defect. In the long-chain approximation, $\gamma_s \approx 0$ and thus the transition is abrupt. As $\gamma$ is further increased, the amplitude on the defect site increases until at $\gamma = \tilde{a} - a$ the bulk energies coalesce in exceptional points and the zero-energy state becomes delocalized. Further on, when the energies of the entire bulk become purely imaginary at $\gamma = \tilde{a} + a$ due to $\mathcal{PT}$-symmetry breaking as will be described in the next subsection, the zero-energy state becomes localized on the defect site itself.

### B. Phase diagrams

Obviously, when the gain parameter $\gamma$ is varied, the entire system goes through several phase transitions. Some of them are unique to the crossed-chain structure. We follow the phases of crossed-chain structures across the parameter ranges.

In both the trivial and topological regimes under the long-chain approximation, when $\gamma^2 \leq (a-\tilde{a})^2$ the bulk energies are purely real, indicating that the bulk is $\mathcal{PT}$-protected, and appear in pairs $(E_m, -E_m)$ for all $m$. As $\gamma$ increases the bulk energies of the states $E_{m,\gamma}$ approach zero as $\sim \pm\sqrt{E_m^2 - \gamma^2}$. When $E_m^2 = \gamma^2$ the bulk states $\pm E_{m,\gamma}$ coalesce at an exceptional point and their energies become a conjugate pair of purely imaginary values, so that $\mathcal{PT}$-symmetry is gradually broken along a series of exceptional points, until the entire bulk is purely imaginary at $\gamma = (a+\tilde{a})^2$. The zero-energy state $E_0$ remains pinned at zero for all values of $\gamma$. We denote the $\mathcal{PT}$-symmetric, -partially-broken and -broken regimes as phases I, II and III, respectively.

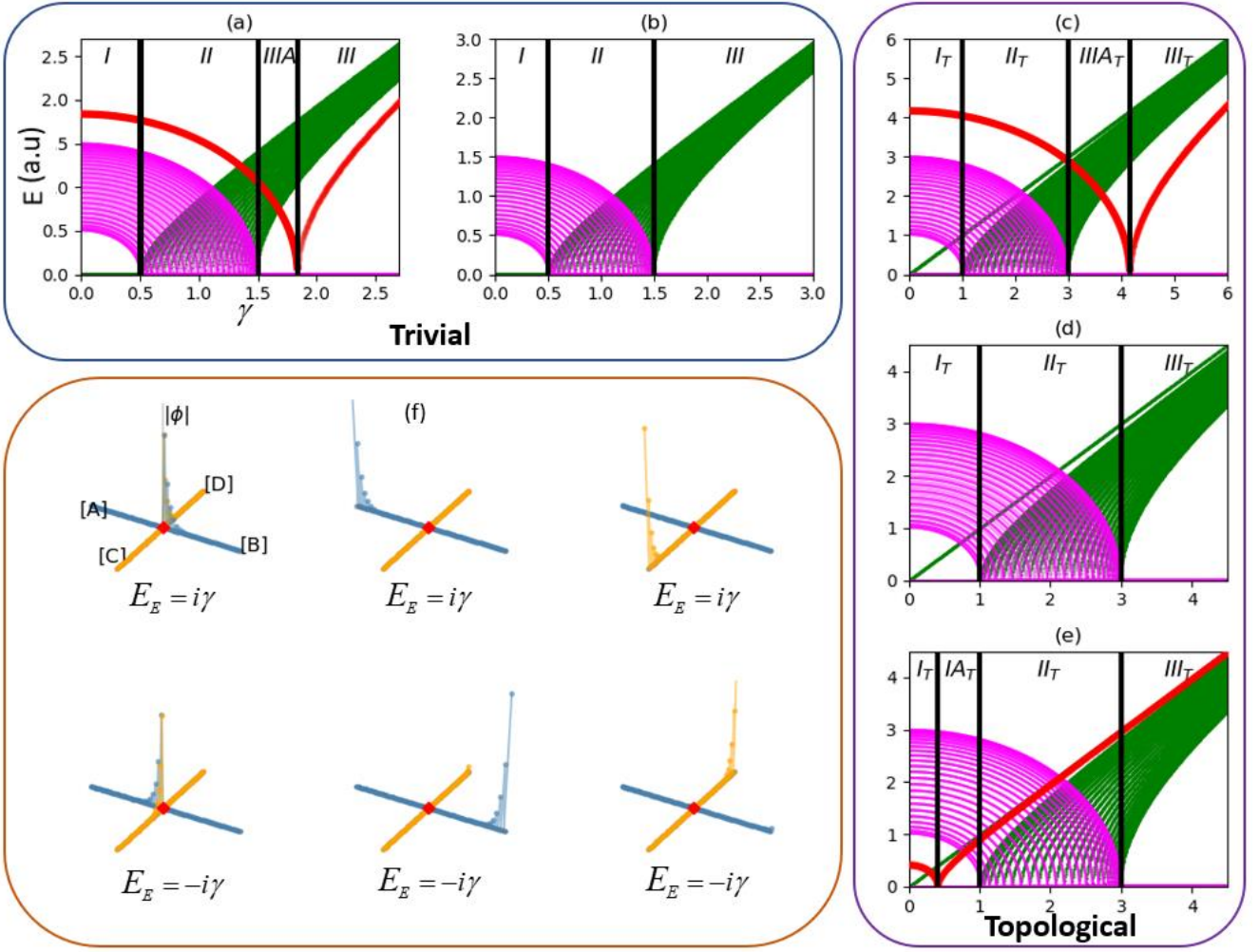

FIG. 6. Phase diagrams for the non-Hermitian crossed chain system as a function of the gain parameter $\gamma$ for $M=50$, for different coupling schemes. Pink and green lines denote the absolute value of the real and imaginary parts, respectively, of the energy pseudospectrum in arbitrary units. Thick red lines denote the isolated localized defect state, when it exists. Phase regions are separated by vertical black lines, and are denoted by roman numerals. (a) Trivial: $\tilde{a}=0.5a$, $\Delta=1.5$ (b) Trivial: $\tilde{a}=0.5a$, $\Delta=0.5$ (c) Topological: $\tilde{a}=2a$, $\Delta=1$ (d) Topological: $\tilde{a}=2a$, $\Delta=0.5$ (e) Topological: $\tilde{a}=2a$, $\Delta=0.12$ (f) Amplitudes of the six edge-states in the topological regime, $\tilde{a}=2a$, $\Delta=1$ and $\gamma=0.5$. The three states on the top have energies $i\gamma$ and the three states on the bottom have energies $-i\gamma$.

In the trivial region, the localized defect state pair $\pm E_\Delta$ exists for $\Delta^2>(\tilde{a}+a)/4\tilde{a}$ (FIG. 6(a)). From Eq. (15), this pair coalesces and becomes purely imaginary at $\gamma=|E_\Delta|$. Above this value, the entire pseudospectrum of the system is purely imaginary and $\mathcal{PT}$-symmetry is broken. However, for $(a+\tilde{a})^2<\gamma<|E_\Delta|$ there exist protected localized states that maintain $\mathcal{PT}$-symmetry, while the entire bulk is $\mathcal{PT}$-broken. This unique phase, denoted IIIA, does not exist in the case of 1D system with a basic defect and is a result of the higher-dimensionality of the crossed-chain system for $\Delta=1$.

For $(\tilde{a}-a)/4\tilde{a}<\Delta^2<(a+\tilde{a})/4\tilde{a}$ (FIG. 6(b)), no localized defect state exists and the $\mathcal{PT}$-symmetry of the entire structure is determined solely from the bulk.

In the topological regime (FIG. 6(c-e)), for $\Delta\neq 0$ there are three degenerate pairs of edge states that, in the long chain approximation, have conjugate imaginary energies (given in Eq. (16)) for $\gamma>0$. These three pairs of states, illustrated in FIG. 6(f), consist of amplitudes localized on the four chain edges on which there is gain (loss), with an energy of $E_E=i\gamma$ ($E_E=-i\gamma$). Therefore, the system is never fully $\mathcal{PT}$-symmetric in the topological regime. We denote the bulk $\mathcal{PT}$-symmetric, -partially-broken and -broken regimes in the topological regime as phases $I_T$, $II_T$ and $III_T$, respectively.

For $\Delta^2>(a+\tilde{a})/4\tilde{a}$ (FIG. 6(c)) there are localized defect states $E_\Delta$ that coalesce at $(a+\tilde{a})^2<\gamma=|E_\Delta|$, above which the entire system is $\mathcal{PT}$-symmetry broken. For

$(a+\tilde{a})^2 < \gamma < |E_\Delta|$ (phase IIIA$_T$), only the localized defect states are $\mathcal{PT}$-symmetric, whereas both the bulk and topological edge states are $\mathcal{PT}$-broken.

For $(\tilde{a}-a)/4\tilde{a} < \Delta^2 < (a+\tilde{a})/4\tilde{a}$ (FIG. 6(d)), no localized defect states exist.

When $\Delta^2 < (\tilde{a}-a)/4\tilde{a}$ (FIG. 6(e)), the localized defect states $E_\Delta$ are inside the band gap, meaning that they coalesce in an exceptional point at $|E_\Delta| = \gamma < (a-\tilde{a})^2$. This creates a unique phase diagram, wherein for $|E_\Delta| < \gamma < (a-\tilde{a})^2$ the bulk pseudospectrum is $\mathcal{PT}$-symmetric while the protected edge and localized defect states are $\mathcal{PT}$-broken. We denote this phase by IA$_T$.

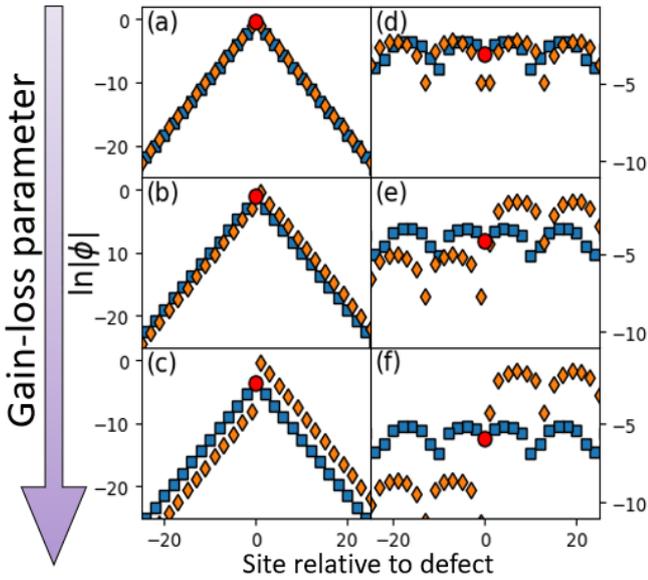

FIG. 7. Amplitude profiles of a localized defect state in the crossed-chain structure along one direction, shown in natural-logarithmic scale, for different values of $\gamma$ for $\tilde{a}=2a$, $\Delta=1$ and $M=50$. Sites are numbered relative to the defect site. The blue squares (orange diamonds) denote the a, c (b, d) sublattices and the red circle denotes the defect site. (a) $\gamma < |E_\Delta|$ - the amplitude profile is spatially symmetric around the defect site. For (b) $\gamma > |E_\Delta|$ and (c) $\gamma \gg |E_\Delta|$: the spatial symmetry of the amplitude profile around the defect breaks; however, the localization length remains identical in subfigures (a)-(c) for all sublattices. Similarly, for an extended singlet state (n=84), at (d) $\gamma < |E_{84}|$ the amplitude profile is symmetric around the defect. For (e) $\gamma > |E_{84}|$ and (f) $\gamma \gg |E_{84}|$: the spatial symmetry breaks, but the wavenumber remains unchanged.

It is important to note that $\mathcal{PT}$-symmetry breaking of any pair of singlet states, is associated with a transition in which the spatial symmetries of the singlet-states amplitude profiles break. We have shown in Appendix F that in the $\mathcal{PT}$-symmetric regime, the amplitude profile of a singlet state is symmetric around the defect site, up to a constant phase shift between the sublattices constituting each SSH chain. This phase shift is equal for one pair of identical chains and has the opposite sign for the complex-conjugate pair of chains (denoted {A, C} and {B, D}, respectively, in FIG. 4). When crossing the exceptional point into the $\mathcal{PT}$-broken regime, two amplitude profiles are obtained – one for every pair of SSH chains. The sublattice amplitudes in different chain pairs are related to one another, and in fact each of the two sublattices in every SSH chain exhibits different amplitudes, but identical wavenumbers, with respect to the other sublattice as shown qualitatively in FIG. 7. To our knowledge, this transition was not discussed in previous works in the context of bulk states, because the division of bulk states into $K-1$ degenerate and spatially-symmetric singlet manifolds is manifested clearly only when a higher-dimensional system with $K>2$ is considered.

We also argue that the non-Hermitian crossed-chain structure illustrated in FIG. 4 is a generalization of the non-Hermitian $\mathcal{PT}$-symmetric 1D structure considered in e.g. [16]. This statement is not trivial, since the crossed-chain description as two coupled SSH pairs is ambiguous. However, the phase transition mentioned above for the higher-dimensionality crossed-chain structure also occurs in the 1D case, as shown in Appendix F. This resemblance demonstrates that coupling two 1D $\mathcal{P}$- and $\mathcal{T}$- (but not $\mathcal{PT}$-) symmetric systems at their center that forms the non-Hermitian crossed-chain system, results in a system exhibiting a behavior of the 1D $\mathcal{T}$- and $\mathcal{PT}$- (but not $\mathcal{P}$-) symmetric configuration.

## V. CONCLUSION

Defect-induced effects on lattices in more than one-dimension have been generally overlooked in the recent spur of research pertaining to topological photonics. In this paper, we have performed a comprehensive analysis of a crossed-chain system, comprised of four identical 1D SSH lattices, coupled at a single point through a mutual site acting as a defect. We have for the first time investigated a topological structure that consists of several SSH chains coupled at a single site with dimension higher than one.

The manifestation of higher dimensionality at the mutual defect site introduces a new type of localized defect states that does not exist for a basic defect in the topologically trivial 1D crossed-chain scheme. Moreover, the higher-dimensionality is one of the parameters that leads to a distinction between the localization lengths of the localized zero-energy state and the defect states in the topologically non-trivial regime.

An important result of our work, is that a Hermitian crossed-chain system consisting of $K$ SSH chains can be decomposed into a reduced representation of $K-1$ identical disconnected SSH chains described by a $K-1$ degenerate manifold, and a single SSH chain connected via a non-reciprocal coupling to an external "reservoir", described by a separate singlet manifold. The former manifold is comprised of states that are superpositions of eigenstates of individual SSH chains (including topological edge states for the topological regime). The latter manifold has spatially symmetric states and a non-zero defect-site amplitude, and includes the localized-defect states. Interestingly, the system describing the SSH chain coupled to an effective reservoir is non-Hermitian. However, both representations of the system are isospectral and so are exactly equivalent.

Our rigorous analysis yields an exact calculation of the energies and wavenumbers of the entire pseudospectrum of the crossed-chain structure as a function of the internal SSH coupling coefficients, as well as the defect coupling strengths. The theory developed here predicts the existence of localized defect states that are isolated from the bulk pseudospectrum and thus robust; furthermore, the energies of these localized states can be tuned by changing the defect coupling strength, and can be placed outside of the bulk, within the energy bandgap, or merged into the bulk states altogether.

These results also apply when non-Hermiticity in the form of alternating on-site gain and loss is introduced to the system. For a system comprised of two pairs of SSH chains as illustrated in FIG. 4, the reduced representation is justified also for the non-Hermitian case. Furthermore, our analysis yielded that the wavenumbers of the non-zero states in the system are not affected by the non-Hermiticity.

Constructing phase diagrams of the crossed-chain system under varying gain-loss parameters, for different topologies and defect coupling strengths, we demonstrated distinctive $\mathcal{PT}$-symmetry regimes for various cases. Some of these phase diagrams have no parallels in 1D with a basic defect, and result from the higher-dimensionality of our proposed structure.

An important contribution is the demonstration of spatial symmetry breaking of the amplitude profile for the singlet states of the system at exceptional points, not only for localized states but also for bulk states. Although this effect was encountered previously for localized defect states in the 1D case, it is due to our analysis of the spatially-symmetric nature of the singlet states (that is manifested in higher-dimensions) that this was proven also for bulk states.

The crossed-chain structure has a potential impact on realizations of coherent arrays of lasers, since it can be used as a building block for a full 2D networked structure with topologically protected lasing sites distributed over the entire 2D area, localized on "defects", rather than the existing proposals of topological coherent lasing only on the 1D interface between 2D topological structures. This paper is also of high relevance for other applications of topological phases, including the study of polymer chains (that was the original seed of the SSH model), where the polymer chains are cross-linked, which is a very typical structure in polymer science.

## APPENDIX A

We re-state the Hamiltonian (3) c.f. (4) in an explicit form for an isotropic crossed-chain configuration (i.e. $M_\sigma = M$, $a_\sigma = a$, $\tilde{a}_\sigma = \tilde{a}$, $\Delta_\sigma = \Delta$). For completeness, we will work in the general case of $K$ identical SSH chains coupled at the defect element. Defining $\psi_n^a = \phi_j$, $j = 2n-1$ and $\psi_n^b = \phi_j$, $j = 2n$ for $1 \le n < M/2$,

$$E_\Delta \psi_n^a = a\psi_n^b + \tilde{a}\psi_{n-1}^b \tag{A1}$$

$$E_\Delta \psi_n^b = a\psi_n^a + \tilde{a}\psi_{n+1}^a \tag{A2}$$

where $E_\Delta$ denotes the defect state energy. Applying the ansatz $\psi_{n-1}^{a/b} = \psi_n^{a/b} \lambda$ for $\lambda = e^{-\kappa}$, $\kappa \in \mathbb{R}^+$ such that $|\lambda| < 1$, Eq. (A1) yields

$$\frac{\psi_n^a}{\psi_n^b} = \frac{a + \tilde{a}\lambda}{E_\Delta}, \; n < M/2, \tag{A3}$$

while plugging Eq. (A2) into Eq. (A1) recovers Eq. (5)

$$E_\Delta^2 = a^2 + \tilde{a}^2 \pm 2a\tilde{a}\cosh\kappa. \tag{A4}$$

The localization length zero-energy state $E_0$ is obtained if the left-hand-side of Eq. (A4) is set to zero and $\kappa$ is replaced by $\eta_0$ (with a minus sign before the hyperbolic cosine so that the LHS vanishes), in agreement with literature for the 1D problem, e.g. Refs. [12,52].

Note that Eq. (5) considers only the plus sign in Eq. (A4), due to the consideration of $\Delta = 1$. For the case $n = M/2$,

$$E_\Delta \psi_{M/2}^b = a\psi_{M/2}^a + \Delta\tilde{a}\psi_\Delta, \tag{A5}$$

where $\psi_\Delta$ denotes the amplitude of the defect state, and

$$E_\Delta \psi_\Delta = K\tilde{a}\Delta\psi_{M/2}^b. \tag{A6}$$

Substituting Eq. (A5) into Eq. (A6) gives us

$$\frac{\psi_M^a}{\psi_M^b} = \frac{1}{a}\left(E_\Delta - \frac{K(\tilde{a}\Delta)^2}{E_\Delta}\right). \tag{A7}$$

Applying the ansatz to Eq. (A7) and using Eq. (A3) one arrives at

$$\lambda = \frac{E_\Delta^2 - K(\tilde{a}\Delta)^2 - a^2}{a\tilde{a}}, \tag{A8}$$

which reaffirms our choice of $\lambda \in \mathbb{R}$. Finally, from plugging the expression for the energy of the defect state Eq. (A4) in Eq. (A8) we arrive at

$$\Delta^2 = \frac{\tilde{a} + a\lambda(s)^{-1}}{K\tilde{a}}, \tag{A9}$$

where $\lambda(s) = e^{is}$, and $s = i\kappa$ if the localized defect-state energy lies outside the bulk pseudospectrum or $s = \pi + i\kappa$ if $E_\Delta$ lies within the gap. By substituting $\Delta = 1$ in Eq. (A9), Eq. (6) from the main text is obtained. Plugging this in back in Eq. (A8),

$$E_\Delta^2 = a^2 + \tilde{a}^2 + \left( \frac{a^2 + (K\Delta^2 - 1)^2 \tilde{a}^2}{K\Delta^2 - 1} \right) \quad (A10)$$

and comparing this to the SSH bulk state expression $E^2(k) = a^2 + \tilde{a}^2 + 2a\tilde{a}\cos k$, we surmise that non-band-gap isolated defect states exist for $\Delta$ satisfying

$$\left| \frac{a^2 + (K\Delta^2 - 1)^2 \tilde{a}^2}{K\Delta^2 - 1} \right| > 2a\tilde{a} + \varepsilon, \quad (A11)$$

where the term $\varepsilon$ stems from the finiteness of the lattice, as the true distribution of the localized states deviates from the ansatz for a finite number of elements. We shall omit $\varepsilon$ in our calculations, as this term is negligible for lattices in the order of 10's of elements which we consider in this paper. Eq. (A11) restates Eq. (A9) under the ansatz of exponential decay and the long-chain approximation

$$\Delta^2 > \frac{a + \tilde{a}}{K\tilde{a}} \quad (A12)$$

for the positive branch of Eq. (A11), or

$$\Delta^2 < \frac{-a + \tilde{a}}{K\tilde{a}} \quad (A13)$$

for its negative branch, implying $\lambda(\kappa) \to \lambda(\kappa - i\pi)$ and leading to the re-statement of the localized defect state energy as

$$E_\Delta^2 = a^2 + \tilde{a}^2 - 2a\tilde{a}\cosh\kappa, \quad \Delta^2 < \frac{\tilde{a} - a}{K\tilde{a}} \quad \text{In gap} \quad (A14)$$

$$E_\Delta^2 = a^2 + \tilde{a}^2 + 2a\tilde{a}\cosh\kappa, \quad \Delta^2 > \frac{a + \tilde{a}}{K\tilde{a}} \quad \text{Outside gap} \quad (A15)$$

with $2a\tilde{a}\cosh(2\kappa) = [a^2 + (K\Delta^2 - 1)^2 \tilde{a}^2]/(K\Delta^2 - 1)$. Note that this expression does not diverge, as $\Delta^2$ is either larger or smaller than $1/K$ when defect states exist, since $\{a, \tilde{a}\} \in \mathbb{R}^+$. Additionally, as seen in Eq. (A15), no localized states exist in the topologically trivial case for $\Delta^2 < \frac{-a + \tilde{a}}{K\tilde{a}}$.

## APPENDIX B

Our goal in this section is to find relations between the energies, wavenumbers and defect coupling strength for the bulk states. We will restrict this calculation to the case of $K = 4$ SSH chains, although generalization to any number can be easily obtained.

Firstly, we formulate an analytical relation between the lattice sites amplitudes in the bulk pseudospectrum of states. For a single SSH chain comprised of $M/2$ unit cells with open boundary conditions, the bulk states will have the general form of sine functions, with wavenumbers $k_m$ quantized for each of the $M$ ($M-2$) bulk states of the trivial (topological) configuration, given by the solution to the eigenvalue problem of the system's Hamiltonian. More specifically, the boundary conditions of the SSH chain read $\psi_0^b = 0$, $\psi_{M/2+1}^a = 0$, which can be restated for the m'th state for the $b$ sublattice as the ansatz

$$\frac{\psi_{n,m}^b}{\sin(k_m n)} = C_m. \quad (B1)$$

for $n \in \{1...M/2\}$ and some constant $C_m$. Defining

$$\chi_{n,m} \equiv \frac{\sin[k_m(n-1)]}{\sin(k_m n)} \quad (B2)$$

and substituting Eq. (B1) in Eqs. (A1) and (A2) would yield the relations

$$\frac{\psi_{n,m}^a}{\psi_{n,m}^b} = \frac{a + \tilde{a}\chi_{n,m}}{E_m}, \quad (B3)$$

$$\frac{\psi_{n+1,m}^a}{\psi_{n,m}^b} = \frac{a/\chi_{n+1,m} + \tilde{a}}{E_m}, \quad (B4)$$

for $\forall n : k_m n \neq 0$. Substituting these back in Eq. (A2) for $n < M/2$ and applying the trigonometric relation $\chi_{n,x} + \chi_{n+1,x}^{-1} = 2\cos x$ yields the known bulk SSH dispersion relation $E^2(k_m) = a^2 + \tilde{a}^2 + 2a\tilde{a}\cos k_m$. Note that this expression does not depend on $n$. Solving for $n = M/2$ and applying the boundary condition $\psi_{M/2+1}^a = 0$ yields the implicit equation for $k_m$

$$E_T^2(k_m) = a^2 + a\tilde{a}\chi_{M/2,m}, \quad (B5)$$

the solutions of which yield the supported wavenumbers for the triplet pseudospectrum.

Secondly, as discussed in the main text, we treat the system as comprised of four degenerate SSH chains connected to an external system with coupling strength $\Delta\tilde{a}$. However, to maintain an identical amplitude profile of the degenerate chains to these in the crossed-chain configuration, we assume that the external system experiences a coupling strength of $4\Delta\tilde{a}$.

We write the equations for the defect and the $n = M/2$ cell,

$$E_S(p)\psi^b_{M/2,p} = \Delta\tilde{a}\psi_{\Delta,p} + a\psi^a_{M/2,p}, \tag{B6}$$

$$E_S(p)\psi^a_{M/2,p} = a\psi^b_{M/2,p} + \tilde{a}\psi^b_{M/2-1,p}, \tag{B7}$$

$$E_S(p)\psi_{\Delta,p} = 4\Delta\tilde{a}\psi^b_{M/2,p}. \tag{B8}$$

The "S" subscript for the energy denotes the energy of the singlet state with the shifted wavenumber $p = k_m + \delta k_m$. Plugging Eqs. (B8) and (B7) in Eq. (B6) for $E_S(p) \neq 0$ yields

$$E_S^2(p) = a^2 + a\tilde{a}\chi_{M/2,p} + 4(\Delta\tilde{a})^2, \tag{B9}$$

which is equivalent to Eq. (B5) when $\Delta = 0$. Substituting the SSH dispersion relation $E^2(p)$ for the shifted wavenumbers $p = k_m + \delta k_m$ in (B9) and reformulating to the implicit form $f(p) = 2\Delta$, we obtain

$$f(p) = \sqrt{1 + \frac{a}{\tilde{a}}\left(\cos(p) + \cot(pM/2)\sin(p)\right)}. \tag{B10}$$

Considering $\Delta > 0$ and solving $f(p) = 2\Delta$ to recover the wavenumbers for a given value of $\Delta$, and substituting the numerically obtained results in the SSH dispersion relation, we recover the eigenvalues of the shifted states obtained from Eq. (3) (FIG. 8). Note that in order to obtain the complex wavenumber of the localized states, one has to substitute $p = i\kappa$ for $\Delta^2 > (a+\tilde{a})/4\tilde{a}$ or $p = \pi + i\kappa$ for $\Delta^2 < (\tilde{a}-a)/4\tilde{a}$ in Eq. (B10).

The wavenumber shifts $\delta k_m$ for the bulk states could be numerically obtained from

$$\delta k_m(\Delta) = p_m(\Delta) - k_m, \tag{B11}$$

where $p_m(\Delta)$ is the m'th element of the ordinal sequence defined by $p(\Delta) \equiv \{p : f(p) = 2\Delta\}$ and $k_m = p_m(0)$. It can be deduced from the cotangent function in Eq. (B10) and also obtained numerically as demonstrated in FIG. 8, that $\max_\Delta |\delta k_m(\Delta)| < |k_m - k_{m+1}|$, so that the shifted singlet energies separate inhomogeneously from the triplet pseudospectrum at $\Delta > 0$, but never exceed the energy level of the next degenerate triplet; taking $\Delta \to \infty$ in $f(p) = 2\Delta$ necessitates $\cot(p_m M) \to \infty$, so that $p_m(\Delta \gg 1) \to \pm m\pi/M$ for $m \in \{1...M/2-1\}$.

## APPENDIX C

Using the same notations as in Appendix A, we start in a similar manner, however – the introduction of gain parameter under $\mathcal{PT}$-symmetry divides the four chains into two pairs – one pair of chains with the loss site adjacent to the defect and

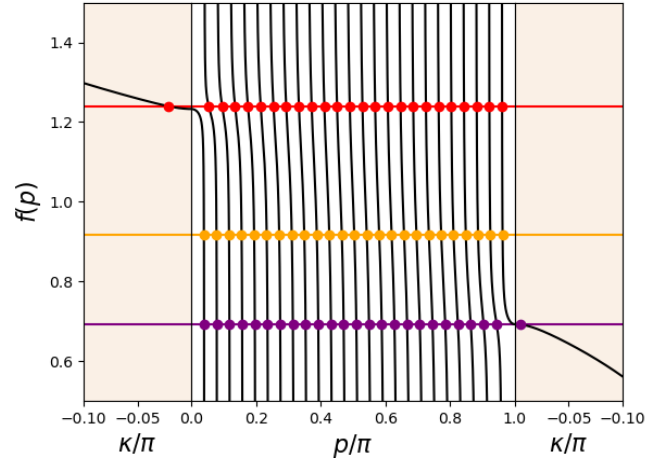

FIG. 8. Numerical calculation of $f(p)$ for $\tilde{a} = 2a$ with $M = 50$. Solutions $p_n(\Delta)$ are the intersections $f(p) = 2\Delta$ marked by full circles. Only $\text{Re}(p) \geq 0$ solutions are presented. The graph is analytically continued for complex wavenumbers $p = i\kappa$ and $p = \pi + i\kappa$ on the left and right panels, respectively, wherein the solutions describe localized states. Red circles (top): $\Delta = \sqrt{(a+\tilde{a})/4\tilde{a}} + 7.5 \cdot 10^{-3}$ with 24 extended and one non-mid-gap solutions (defect states are excluded from the bulk and are localized), orange circles (middle): $\Delta = 0.75\sqrt{(a+\tilde{a})/4\tilde{a}}$ with 25 extended solutions (defect states are merged into the bulk), purple circles (bottom): $\Delta = \sqrt{(\tilde{a}-a)/4\tilde{a}} - 7.5 \cdot 10^{-3}$ with 24 extended and one localized in-gap solutions (defect states are excluded from the bulk and are localized).

one pair with the gain site adjacent to it (FIG. 9). We will focus on the crossed-chain case ($K = 4$). Generalizations of these calculations can be done for an even number of SSH chains, as our assumptions rely on $\mathcal{PT}$-symmetry in the system. Without loss of generality,

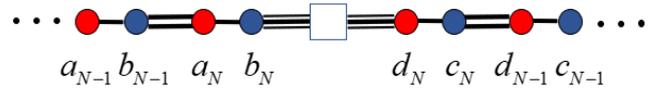

FIG. 9. Schematic of one axis of the complex-SSH crossed-chain configuration. The white square denotes the defect site, red circles denote gain $\gamma$, blue circles denote loss $-\gamma$, black line denotes a coupling strength $a$, double-line denotes a coupling strength $\tilde{a}$ and finally triple-line denotes the defect coupling strength $\tilde{a}\Delta$.

$$(E_{\Delta,\gamma} - i\gamma)\psi^a_n = a\psi^b_n + \tilde{a}\psi^b_{n-1} \tag{C1}$$

$$(E_{\Delta,\gamma} + i\gamma)\psi^b_n = a\psi^a_n + \tilde{a}\psi^a_{n+1} \tag{C2}$$

for the first pair of chains and

$$\left(E_{\Delta,\gamma}+i\gamma\right)\psi_n^c = a\psi_n^d + \tilde{a}\psi_{n-1}^d \tag{C3}$$

$$\left(E_{\Delta,\gamma}-i\gamma\right)\psi_n^d = a\psi_n^c + \tilde{a}\psi_{n+1}^c \tag{C4}$$

for the second, for $n < M/2$.

Assuming again that $\psi_{n-1}^x = \psi_n^x \lambda$ with $x \in \{a,b,c,d\}$, $\lambda = |\lambda|e^{i\varphi}$, $|\lambda|<1$ and some phase $\varphi \in \mathbb{R}$ for the localized non-mid-gap defect state, plugging Eq. (C2) in Eq. (C1) recovers

$$E_{\Delta,\gamma}^2 = a^2 + \tilde{a}^2 + a\tilde{a}\left(\lambda+\lambda^{-1}\right) - \gamma^2, \tag{C5}$$

which suggests that $E_{\Delta,\gamma}$ is either a purely real or a purely imaginary quantity, depending on the magnitude of $\gamma$. Replacing $\lambda = e^{-\kappa}$ or $\lambda = e^{i\pi-\kappa}$ restores Eqs. (15) or (16), respectively. We shall first consider $E_{\Delta,\gamma} \in \mathbb{R}$. Since applying the complex conjugate operator to Eqs. (C3) and (C4) yields Eqs. (C1) and (C2), respectively, it is clear that $\psi_n^a = \psi_n^{c*}$ and $\psi_n^b = \psi_n^{d*}$, so that Eqs. (C4) and (C3) become redundant. We write the equation for the defect site

$$E_{\Delta,\gamma}\psi_\Delta = 2\tilde{a}\Delta\psi_{M/2}^b + 2\tilde{a}\Delta\psi_{M/2}^{b*} \tag{C6}$$

and apply the ansatz, and since $\psi_\Delta \in \mathbb{R}$ we can conclude that $\varphi = 0$ and so $\lambda \in \mathbb{R}$. We extract from (C1) the relation

$$\frac{\psi_n^a}{\psi_n^b} = E_{\Delta,\gamma}\frac{a+\tilde{a}\lambda}{E_{\Delta,\gamma}^2+\gamma^2} + i\gamma\frac{a+\tilde{a}\lambda}{E_{\Delta,\gamma}^2+\gamma^2}, \tag{C7}$$

for $n < M/2$, from which we obtain

$$\operatorname{Im}\psi_n^a = \frac{a+\tilde{a}\lambda}{E_{\Delta,\gamma}^2+\gamma^2}\left(E_\Delta \operatorname{Im}\psi_n^b + \gamma \operatorname{Re}\psi_n^b\right), \tag{C8}$$

$$\operatorname{Re}\psi_n^a = \frac{a+\tilde{a}\lambda}{E_{\Delta,\gamma}^2+\gamma^2}\left(E_\Delta \operatorname{Re}\psi_n^b - \gamma \operatorname{Im}\psi_n^b\right). \tag{C9}$$

For $n = M/2$, we write

$$\left(E_{\Delta,\gamma}+i\gamma\right)\psi_{M/2}^b = a\psi_{M/2}^a + \Delta\tilde{a}\psi_\Delta. \tag{C10}$$

Solving separately for the real and imaginary parts of Eq. (C10),

$$E_{\Delta,\gamma}\operatorname{Re}\psi_{M/2}^b - \gamma\operatorname{Im}\psi_{M/2}^b = a\operatorname{Re}\psi_{M/2}^a + \Delta\tilde{a}\operatorname{Re}\psi_\Delta \tag{C11}$$

$$E_{\Delta,\gamma}\operatorname{Im}\psi_{M/2}^b + \gamma\operatorname{Re}\psi_{M/2}^b = a\operatorname{Im}\psi_{M/2}^a + \Delta\tilde{a}\operatorname{Im}\psi_\Delta. \tag{C12}$$

Since $\operatorname{Im}\psi_\Delta = 0$, multiplying Eq. (C12) by $\lambda^{M/2-n}$ and plugging the result in Eq. (C8) yields

$$\operatorname{Im}\psi_n^a = \frac{a+\tilde{a}\lambda}{E_{\Delta,\gamma}^2+\gamma^2}\left(a\operatorname{Im}\psi_n^a\right), \tag{C13}$$

which for $\tilde{a} \neq 0$ can be true iff $\forall n: \operatorname{Im}\psi_n^a \equiv 0$. Taking Eqs. (C11) and (C12) c.f. Eq. (C6),

$$\frac{\psi_{M/2}^b}{\psi_{M/2}^a} = a\frac{E_{\Delta,\gamma}+i\gamma}{E_{\Delta,\gamma}^2+\gamma^2-4(\tilde{a}\Delta)^2}. \tag{C14}$$

Comparing this with Eq. (C7) results in

$$\lambda = \frac{E_{\Delta,\gamma}^2+\gamma^2-4(\tilde{a}\Delta)^2-a^2}{a\tilde{a}}, \tag{C15}$$

and using Eq. (C5) one has

$$\Delta^2 = \frac{\tilde{a}+a\lambda^{-1}}{4\tilde{a}}. \tag{C16}$$

This reproduces Eq.(10) from the main text, suggesting that the defect state remains unaffected by the introduction of on-site non-Hermiticity in the considered scheme while $E_{\Delta,\gamma} \in \mathbb{R}$. Furthermore, the non-mid-gap defect state energies Eq. (C5) can be re-expressed to yield a generalization of Eq. (A10),

$$E_{\Delta,\gamma}^2 = a^2 + \tilde{a}^2 + \left(\frac{a^2+\left(4\Delta^2-1\right)^2\tilde{a}^2}{4\Delta^2-1}\right) - \gamma^2. \tag{C17}$$

A transition of the system occurs at the $\mathcal{PT}$-symmetry breaking point $E_{\Delta,\gamma}^2 = 0$, beyond which point $E_{\Delta,\gamma}$ becomes purely imaginary, and therefore the relations between Eqs. (C1) and (C2) to Eqs. (C3) and (C4) no longer imply that the two chain pairs defined by their lattice sites $\{a,b\}$ and $\{c,d\}$, respectively, are complex conjugates. Thus, we must assume two dissimilar pairs of complex SSH chains. Denoting $E_{\Delta,\gamma} = iW$ with $W \in \mathbb{R}$ for convenience, we have

$$iW\psi_\Delta = 2\tilde{a}\Delta\psi_{M/2}^b + 2\tilde{a}\Delta\psi_{M/2}^d, \tag{C18}$$

$$i\left(W-\gamma_\mu\right)\psi_{M/2}^\mu = a\psi_{M/2}^\nu + \tilde{a}\Delta\psi_\Delta, \tag{C19}$$

where we have defined $\mu \in \{b,d\}$ and $\nu \in \{a,c\}$ such that the pair $(\mu,\nu) \in \{(b,a),(d,c)\}$. Additionally, $\gamma_b = -\gamma_d = -\gamma$. We keep the assumption of a localized state, wherein the localization length remains dependent only on the coupling coefficients of the structure. To justify this for dissimilar chains, we denote the evanescence factors $\lambda_{ab}$ and $\lambda_{cd}$, and plug Eq. (C2) into Eq. (C1) and Eq. (C4) into Eq. (C3), recovering Eq. (C5) respectively for $E_{\Delta,\gamma}^2(\lambda_{ab},\gamma)$ and $E^2_{\Delta,\gamma}(\lambda_{cd},\gamma)$. Equating these expressions yields a true statement if and only if $\lambda_{ab} = \lambda_{cd}$.

Denoting $\lambda \equiv \lambda_{ab} = \lambda_{cd}$, from Eqs. (C1) and (C3) we obtain

$$\frac{\psi^\nu_n}{\psi^\mu_n} = -i\frac{a+\tilde{a}\lambda}{W+\gamma_\mu}, \tag{C20}$$

for $1 \leq n < M/2$. It is implied from Eq. (C20) that the elements in each unit cell are shifted by a constant phase of $\pi/2$ with respect to one another. Solving Eq. (C19) c.f. Eq. (C20) separately for the real and imaginary parts and consolidating the expressions, we arrive at

$$\psi^\mu_{M/2} = -i\frac{\tilde{a}\Delta(W+\gamma_\mu)}{W^2-\gamma^2+a^2+a\tilde{a}\lambda}\psi_\Delta. \tag{C21}$$

Plugging these back in Eq. (C19) generalizes Eq. (C20) to $n = M/2$. Finally,

$$\psi^a_{M/2} = \frac{-\tilde{a}\Delta(a+\tilde{a}\lambda)}{W^2-\gamma^2+a^2+a\tilde{a}\lambda}\psi_\Delta, \tag{C22}$$

$$\psi^c_{M/2} = \frac{-\tilde{a}\Delta(a+\tilde{a}\lambda)}{W^2-\gamma^2+a^2+a\tilde{a}\lambda}\psi_\Delta, \tag{C23}$$

so that

$$\psi^a_n = \psi^c_n, \tag{C24}$$

$$\psi^b_n = \frac{W-\gamma}{W+\gamma}\psi^d_n \tag{C25}$$

for all $n$. This result demonstrates a breaking of spatial symmetry in the system that occurs when the state crosses an exceptional point and $\mathcal{PT}$-symmetry breaks, while the symmetric inter-cell exponential decay ansatz (see FIG. 7 in the main text) is maintained. When the gain parameter is very large, i.e. $\gamma \gg 1$ (so that $\gamma \approx -W$), the site amplitudes of the $\{a,b\}$ chains are held at zero, and the site with the highest amplitude is the $\psi^d_{M/2}$ site, as $|\psi^\Delta| \to 0$. This effectively demonstrates a negligible penetration of energy into the defect and the $\{a,b\}$ chains. From Eqs. (C18) and (C21)-(C23) we recover Eq. (C16), reaffirming that the localization length depends on the defect strength of the structure and not on its gain parameter.

## APPENDIX D

We combine the assumptions of Appendix B and Appendix C to analyze the relations between the energies of the bulk states, the defect coupling strength and the gain parameter.

We start by applying the same ansatz (B1), since the addition of a constant imaginary gain (loss) term to every second site should not affect our underlying assumptions, as one can define a complex energy term $\varepsilon_q \equiv E_q + (-)i\gamma$ for the sublattices of the sites adjacent to the defect. Since now we have two "types" of SSH chains, as illustrated in FIG. 9, we re-state the ansatz (B1)

$$\frac{\psi^\mu_{n,m}}{\sin(k_m n)} = C^\mu_m \tag{D1}$$

With the same $(\mu,\nu)$ notation of Appendix C. Plugging Eq. (D1) into Eqs. (C1)-(C4) with the bulk states energies denoted $E_\gamma(k_m)$ instead of $E_{\Delta,\gamma}$ and the definition (B2) yields

$$\frac{\psi^\nu_{n,m}}{\psi^\mu_{n,m}} = \frac{a+\tilde{a}\chi_{n,m}}{E_\gamma(k_m)+i\gamma_\mu}, \tag{D2}$$

$$\frac{\psi^\nu_{n+1,m}}{\psi^\mu_{n,m}} = \frac{a/\chi_{n+1,m}+\tilde{a}}{E_\gamma(k_m)-i\gamma_\mu} \tag{D3}$$

for $\forall n < M/2 : k_m n \neq 0$. From substitution of Eqs. (D2) and (D3) in Eq. (C2) we recover the dispersion relation $E^2_m(k_m) = a^2 + \tilde{a}^2 + 2a\tilde{a}\cos k_m - \gamma^2$. The triplet energies $E_{T,\gamma}(k_m)$ are obtained by imposing the boundary condition $\psi^\nu_{M/2+1,m} = 0$ in the same manner as in Appendix B and we get

$$E^2_{T,\gamma}(k_m) + \gamma^2 = a^2 + a\tilde{a}\chi_{M/2,m}. \tag{D4}$$

For the singlet bulk energies $E_{S,\gamma}(p)$, we note that $E_{S,\gamma}(p)$ can be either purely real or purely imaginary. We begin with the case $E_{S,\gamma}(p) \in \mathbb{R}$, for which $\psi^b_{n,m} = \psi^{d*}_{n,m}$ for all $n$ and $m$. We write the equations for the effective system that describes the singlet states,

$$(E_{S,\gamma}(p)-i\gamma_\mu)\psi^\mu_{M/2,p} = \Delta\tilde{a}\psi_{\Delta,p} + a\psi^\nu_{M/2,p}, \tag{D5}$$

$$E_{S,\gamma}(p)\psi_{\Delta,p} = 4\Delta\tilde{a}\,\mathrm{Re}\,\psi^\mu_{M/2,p}, \tag{D6}$$

and we solve Eq. (D5) separately for the real and imaginary parts of $\psi^\mu_{M/2,p}$. Plugging in Eq. (D6), we obtain for $E_{S,\gamma} \neq 0$

$$\frac{\psi^\nu_{M/2,p}}{\psi^\mu_{M/2,p}} = \frac{1}{a}\frac{E^2_{S,\gamma}(p)+\gamma^2-4(\tilde{a}\Delta)^2}{E^2_{S,\gamma}(p)+\gamma^2}(E_{S,\gamma}(p)+i\gamma_\mu) \tag{D7}$$

which together with Eq. (D2) yields

$$E^2_{S,\gamma}(p)+\gamma^2 = a^2 + \tilde{a}a\chi_{M,q} + 4(\tilde{a}\Delta)^2. \tag{D8}$$

Since in the non-Hermitian case $E^2_{S,\gamma}(p) = a^2 + \tilde{a}^2 + 2a\tilde{a}\cos p - \gamma^2$, this yields exactly Eq. (B9), and so the solution of $f(p) = 2\Delta$ for $\Delta > 0$ and $f(p)$ as defined in Eq. (B10) generates the same solutions $p_m(\Delta)$ as in the Hermitian case. The case of $E_{s,\gamma} = 0$ recovers the $E_0$ state. Turning to consider $E_q \in i\mathbb{R}$, we define $W(p) \in \mathbb{R}$ through $E_s(p) = iW(p)$ and re-write the equation for the defect site

$$iW(p)\psi_{\Delta,p} = 2\Delta\tilde{a}\psi^b_{M/2,p} + 2\Delta\tilde{a}\psi^d_{M/2,p}. \tag{D9}$$

Solving Eq. (D5) for its real and imaginary parts along with Eq. (D2) once for $\{a,b\}$ and once for $\{c,d\}$ results in

$$\frac{\psi^\mu_{M/2,p}}{\psi_{\Delta,p}} = -i\frac{\tilde{a}\Delta(W(p)+\gamma_\mu)}{W^2(p)-\gamma^2+a^2+a\tilde{a}\chi_{M/2,p}}, \tag{D10}$$

implying from Eq. (D2) that as in the case of localized defect states, when $E_s(p) \in i\mathbb{R}$ - although the two SSH pairs amplitude profiles are not spatially symmetric around the defect as in the Hermitian case - $\psi^c_{n,p} = \psi^a_{n,p}$ for all $n$ and $p$. In fact, it can be readily obtained that Eqs. (C22)-(C25) apply here under the replacement of $\lambda$ with $\chi_{M/2,p}$ and adequate notations.

Combining Eqs. (D9) and (D10) results in

$$W^2(p) + \gamma^2 = a^2 + a\tilde{a}\chi_{M/2,p} + 4(\tilde{a}\Delta)^2, \tag{D11}$$

which again yields the exact expression as Eq. (B9), for $E_S^2(p) < 0$. Therefore, we conclude that the solution for the wavenumber equation is not affected by the introduction of non-Hermiticity in the scheme suggested in this paper, and that the relation between the defect coupling strength and the shifts in wavenumbers is constant for all $\gamma$.

## APPENDIX E

In this appendix we will succinctly establish several properties of the localized states of 1D crossed-chain system in the topological regime, with a basic defect ($\Delta = 1$), since to our knowledge this case was not thoroughly addressed in literature.

First, we will demonstrate the phase transition that the zero state $E_0$ undergoes between the trivial and topological regimes. In FIG. 10 we show the amplitude profile of a 1D crossed-chain system in the trivial and topological regimes, as well as in the singular transition point $\tilde{a} = a$. In the trivial regime (stronger inter-coupling), only the defect and the A sublattices have non-zero amplitudes, and the amplitude profile is exponentially decreasing from the defect site. As $\tilde{a}$ increases in the trivial regime, the band gap shrinks and the localization length increases, until the gap closes at $\tilde{a} = a$ and the zero-energy state $E_0$ becomes delocalized. When $\tilde{a} > a$, the gap re-opens in the topological regime and the localization shifts to the outer edges of the A sublattices, as now the inter-coupling is stronger and so the defect is strongly coupled to its neighbouring sites, so the state "diffuses" from it rather than center around it. The amplitudes of the B lattice remain pinned at zero, as can be seen from the Hamiltonian equations of the system for $E = 0$ when a non-zero amplitude at the A lattice is assumed.

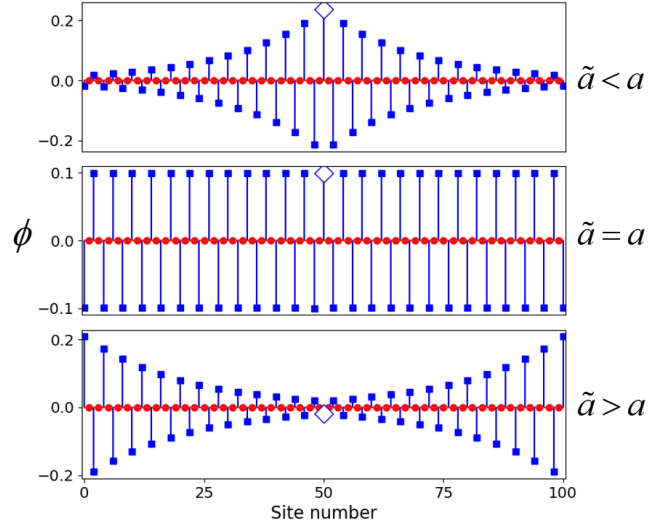

FIG. 10. Zero-energy state of a 1D crossed chain for a trivial SSH (top), a transition point SSH (mid) and a topological SSH (bottom) lattices of length $M = 50$. Blue squares indicate sublattice A, red circles indicate sublattice B and the blue diamond indicates the defect site.

The localization length in the topological region is then given by $\eta_0^{topo} = -\log(a/\tilde{a})$, and so the general expression of the localization length of $E_0$ is

$$\eta_0 = |\log(a/\tilde{a})|. \tag{E1}$$

The pseudospectrum of the system is illustrated in FIG. 11 for the parameters $\tilde{a} = 1.75a$ and $M = 50$. The red diamonds in the insets (a.1) and (a.2) depict the edge states $\pm E_\Delta$, whose properties can be calculated from Appendix A for a basic defect and $K = 2$ in the long-lattice approximation,

$$E_\Delta^2 = a^2 + \tilde{a}^2 + 2a\tilde{a}\cosh\kappa, \tag{E2}$$

$$\kappa = -\log(a/\tilde{a}). \tag{E3}$$

The three states in inset (a.3) are the zero-energy state $E_0$ in the center, and two topological SSH edge states $E_E$ on each site. The energies and localization lengths of these states are known [74], and in the long-lattice approximation are

$$E_E \approx 0, \tag{E4}$$

which is true if and only if

$$\eta_E \approx -\log(a/\tilde{a}). \tag{E5}$$

The long-lattice approximation can be justified for our parameters as seen in FIG. 11(a.3), wherein the scale of the energy axis demonstrates the proximity of $E_E$ to zero.

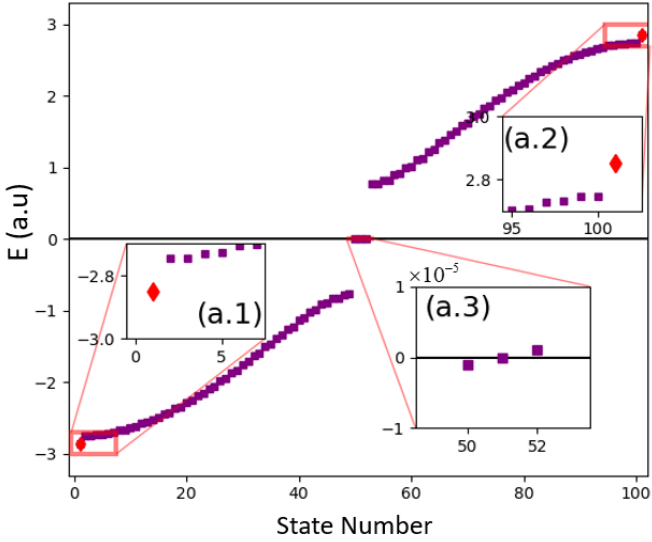

FIG. 11. Pseudospectrum of the 1D SSH crossed chain in the topological regime, with $\tilde{a} = 1.75a$ and $M = 50$. Insets (a.1) and (a.2) depict the localized defect states $E_\Delta$ (red diamonds), and inset (a.3) depicts the zero-energy state $E_0$ at the center and the two topological SSH edge-states around it. The scale of the energy axis in inset (a.3) demonstrates the precision of the long-chain approximation for the considered system parameters.

## APPENDIX F

We briefly present an explanation for the behavior of the amplitude profile of a non-Hermitian crossed SSH system with an even number of chains, in the $\mathcal{PT}$-symmetry-preserved and -broken regimes to support our claim in the main text. This explanation is also valid for a 1D crossed-chain.

While $\mathcal{PT}$-symmetry is maintained, only the relative phases between the sublattices of each SSH chain depend on $\gamma$, while the amplitudes ratio remains constant. Eqs. (C7) and (D2) that describe the intra-cell amplitude ratios have the general form

$$\frac{\psi_{n,m}^{a/c}}{\psi_{n,m}^{b/d}} = \frac{a + \tilde{a}z}{E(\gamma) \pm i\gamma} \equiv \frac{q}{g(\gamma)}, \tag{F1}$$

with $z \in \mathbb{R}$ taking the role of the inter-cell amplitude ratio $\lambda$ or $\chi_{m,n}$ for a localized or extended state, respectively. The above is true for a system being comprised of SSH chains and their complex conjugates, thus consisting of four sublattices denoted $\{a,b,c,d\}$, and therefore can be also applied to the 1D case. The plus or minus sign depends on by whether the sublattice $a/c$ has on-site gain or loss, respectively. Considering the dependence of $E(\gamma)$ on $\gamma$, Eq. (F1) has a real numerator that does not depend on $\gamma$ and a complex denominator of the form $g(\gamma) = [C^2 - \gamma^2]^{1/2} \pm i\gamma$, $C \in \mathbb{R}^+$. While $|\gamma| \leq C$,

$$|g(\gamma)| = C, \quad \angle g(\gamma) = \pm\arctan\left(\gamma / [C^2 - \gamma^2]^{1/2}\right), \tag{F2}$$

whereas for $|\gamma| > C$

$$|g(\gamma)| = |[\gamma^2 - C^2]^{1/2} \pm \gamma|, \quad \angle g(\gamma) = \pm\pi/2. \tag{F3}$$

Eq. (F2) shows that in the $\mathcal{PT}$-symmetric regime, $|\angle g(\gamma)|$ monotonically increases from 0 for $\gamma = 0$, to $\pi/2$ at $|\gamma| = C$. When the $\mathcal{PT}$-symmetry is broken (Eq. (F3)), the sublattices maintain a constant relative $\pi/2$ phase while their amplitudes ratio change, since in this region $|g(\gamma)|$ changes monotonically with $\gamma$ (depending on the sign $\text{Im}\, g(\gamma)$), while its phase remains constant. This demonstrates the separation of the amplitudes of the sublattices in a single SSH chain when $\mathcal{PT}$-symmetry breaks. When two mirrored SSH lattices are coupled as in FIG. 9, such that the imaginary on-site potentials of the sublattices adjacent to the defect have opposite signs, the sublattice amplitudes ratios in each chain are described by Eq. (F3) with opposing signs. This proves that the amplitude symmetry of the non-Hermitian crossed-SSH system breaks when $\mathcal{PT}$-symmetry is broken, for 1D as well as for higher dimension crossed-chain systems, with the gain-loss scheme considered in the paper.

## APPENDIX G

As an indication for the topological phase in the non-Hermitian regime, we numerically demonstrate the robustness of the topological states of the system against structural disorder, following [13,16] and others.

We do so by considering $\mathcal{PT}$-symmetry- and chirality-preserving disorder of the coupling coefficients, so that the coupling coefficients of the n'th cell are

$$a_n = a + S(\tilde{a} - a)\frac{U_n}{2}, \tag{G1}$$

$$\tilde{a}_n = \tilde{a} - S(\tilde{a} - a)\frac{U_n}{2}, \tag{G2}$$

with $U_n$ a random number from a uniform distribution in the range $[-1,1]$ and $S \in \mathbb{R}$ the disorder strength parameter. These disordered coupling coefficients are identical in all four

SSH chains, so that $\mathcal{PT}$-symmetry is maintained for any realization as long as $S \leq 1 - \gamma/|\tilde{a}-a|$. The amplitudes of the topological zero, defect and edge states on one of the axes of the crossed-chain structure are shown in FIG. 12 for 300 realizations of disorder, for $S=0.5$, $\tilde{a}=2a$ and $\gamma=0.5$. Note that this is the largest disorder strength parameter for the considered coupling and gain-loss scheme according to the condition above for which $\mathcal{PT}$-symmetry is conserved.

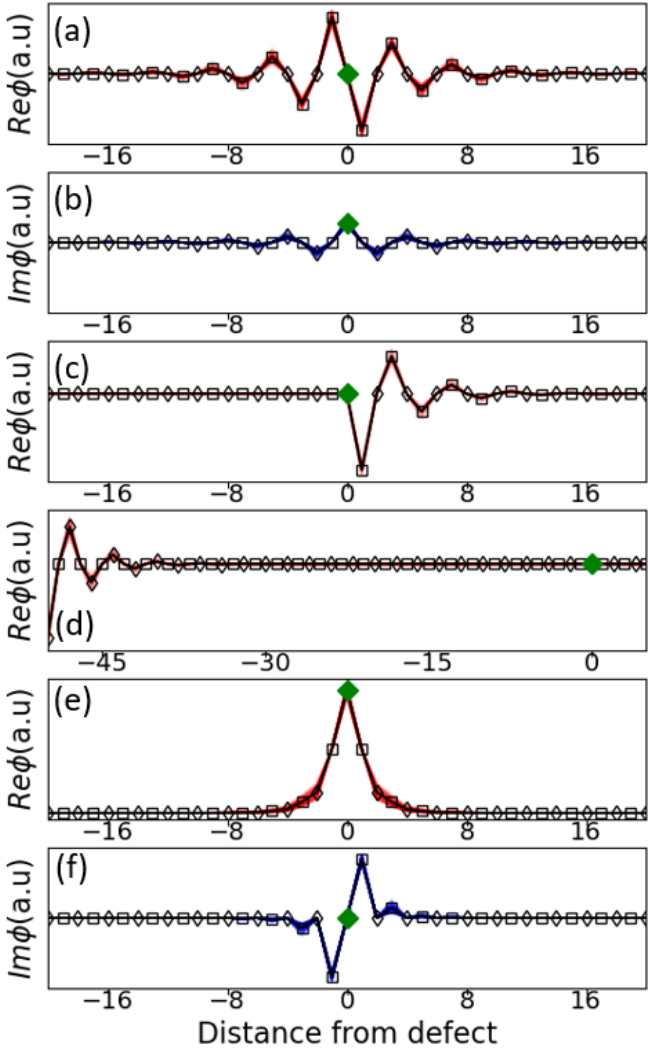

FIG. 12. Amplitude profiles of localized states in a non-Hermitian crossed-chain SSH structure with $\gamma=0.5$, $\tilde{a}=2a$, $M=50$ along a single axis. Hollow diamonds mark sublattice A/C and hollow squares mark sublattice B/D. Full green diamond marks the defect site amplitude. Each subfigure portrays 300 realizations of random disorder with $S=0.5$ in red (real) or blue (imaginary) lines. Black lines denote the average amplitude profile. (a) and (b) are, respectively, the real and imaginary part of the zero state. (c) and (d) are, respectively, inner and outer edge states. (e) and (f) are, respectively, the real and imaginary parts of a localized defect state.

It is evident from FIG. 12 that the amplitudes do not change appreciably when disorder is introduced. To quantify that, we have numerically calculated the imaginary wavenumbers in each realization by exponential fitting of the amplitude profile envelope and performed an analysis on their distributions. For the zero state we found the mean value to its distribution be $\text{Im}\,\bar{\eta}_0 \approx 0.603$ with variance $\sigma^2_{\eta_0} \approx 8.657 \cdot 10^{-3}$ ($100 \times \sigma^2_{\eta_0}/\text{Im}\,\bar{\eta}_0 \approx 1.4\%$), for the edge states $\text{Im}\,\bar{\eta} \approx 0.694$ with $\sigma^2_{\eta} \approx 7.012 \cdot 10^{-3}$ ($100 \times \sigma^2_{\eta}/\text{Im}\,\bar{\eta} \approx 1\%$) and for the defect states $\text{Im}\,\bar{\kappa} \approx 1.794$ with $\sigma^2_{\kappa} \approx 28.29 \cdot 10^{-3}$ ($100 \times \sigma^2_{\kappa}/\text{Im}\,\bar{\kappa} \approx 1.6\%$). The very small variances around the respective means suggest robustness against this large disorder. Similar variances of the wavenumbers in different realizations are also obtained in the Hermitian case, $\gamma=0$. Additionally, all the means converge to their zero-disorder values. This robustness demonstrates the topological nature of the crossed-chain structure also in passing to the non-Hermitian regime as discussed in the main text.

Further indication of the topology is the persistence of the property of the zero-state of having purely real amplitudes on the B/D sublattices and purely imaginary amplitudes on the A/C sublattices, with any disorder strength parameter in the above range. These numerically-calculated mean square errors (MSE) for the real and imaginary parts of sublattices A/C and B/D, with respect to an ordered amplitude profile $\phi_0$, are shown in FIG. 13. We have defined the MSE between two vectors of length $n$ as $MSE(\vec{x},\vec{y}) = \sum_n \left| x_n^2 - y_n^2 \right|^2$.

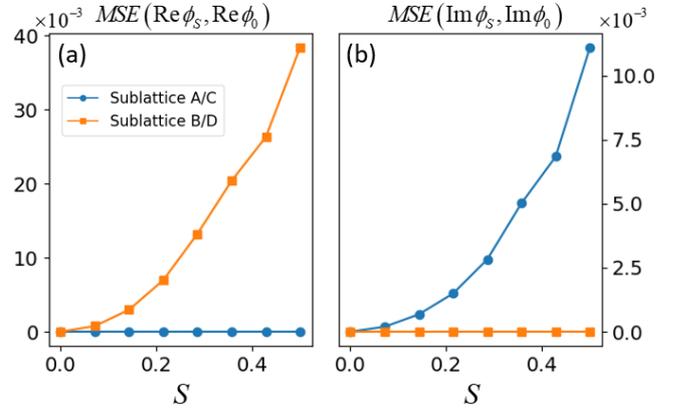

FIG. 13. Effect of the disorder strength coefficient $S$ on the real and imaginary parts of the zero state amplitude profiles in sublattices A/C and B/D in a non-Hermitian crossed-chain SSH structure with $\gamma=0.5$, $\tilde{a}=2a$, $M=50$. The circles (squares) denote the mean value of the MSE for sublattice A (B) over 300 realizations of the disorder. (a) MSE of the real part of the zero-state amplitude. (b) MSE of the imaginary part of the zero-state amplitude.